\documentclass[11pt,letterpaper]{article}

\usepackage[utf8]{inputenc}
\usepackage[T1]{fontenc}
\usepackage{amsmath,amssymb,amsfonts,amsthm,mathtools}
\IfFileExists{libertine.sty}{\usepackage[tt=false,type1=true]{libertine}}{\usepackage{lmodern}}
\IfFileExists{zi4.sty}{\usepackage[varqu]{zi4}}{}
\IfFileExists{libertine.sty}{\IfFileExists{newtxmath.sty}{\usepackage[libertine]{newtxmath}}{}}{\IfFileExists{newtxmath.sty}{\usepackage{newtxmath}}{}}
\usepackage[margin=1in]{geometry}
\usepackage{microtype,parskip,needspace,placeins}
\usepackage{enumitem,booktabs,array,float,subcaption}
\usepackage[ruled,vlined,linesnumbered,norelsize]{algorithm2e}
\usepackage[dvipsnames]{xcolor}
\usepackage{tikz}
\usetikzlibrary{arrows.meta,shapes.geometric,positioning,fit,backgrounds,calc,decorations.pathreplacing}
\definecolor{ForestGreen}{rgb}{0.1333,0.5451,0.1333}
\definecolor{DarkRed}{rgb}{0.65,0,0}
\usepackage{hyperref}
\hypersetup{colorlinks=true,linkcolor=DarkRed,urlcolor=cyan,citecolor=ForestGreen,
 pdftitle={Nonuniform Buy-at-Bulk Network Design: Improved Approximation and LP Integrality Gap},
 pdfauthor={Iman Gholami, MohammadTaghi Hajiaghayi}}
\usepackage[nameinlink]{cleveref}
\newtheorem{theorem}{Theorem}
\newtheorem{corollary}{Corollary}[section]
\newtheorem{lemma}[corollary]{Lemma}
\newtheorem{proposition}[corollary]{Proposition}
\theoremstyle{definition}
\newtheorem{definition}[corollary]{Definition}
\theoremstyle{remark}

\newcommand{\numberthis}{\refstepcounter{equation}\tag{\theequation}}
\newcommand{\LP}{\operatorname{LP}}

\newcommand{\OPT}{\operatorname{OPT}}
\DeclareRobustCommand{\FullRef}[2]{\hyperref[#2]{#1}}

\newcommand{\E}{\mathbb E}

\tikzset{
  triangle/.style args={#1}{
    insert path={
        ++(90:1.2*#1)
      -- ++($(90:-1.2*#1)+(210:1.2*#1)$)
      -- ++($(210:-1.2*#1)+(330:1.2*#1)$)
      -- cycle
    }
  },
  pentagon/.style args={#1}{
    insert path={
      ++(90:1.05*#1)
      -- ++($(90:-1.05*#1)+(162:1.05*#1)$)
      -- ++($(162:-1.05*#1)+(234:1.05*#1)$)
      -- ++($(234:-1.05*#1)+(306:1.05*#1)$)
      -- ++($(306:-1.05*#1)+(18:1.05*#1)$)
      -- cycle
    }
  },
  square/.style args={#1}{
    insert path={
        ++(45:1.1*#1)
      -- ++($(45:-1.1*#1)+(135:1.1*#1)$)
      -- ++($(135:-1.1*#1)+(225:1.1*#1)$)
      -- ++($(225:-1.1*#1)+(315:1.1*#1)$)
      -- cycle
    }
  },
  star/.style args={#1}{
    insert path={
      ++(90:1.1*#1)
      \foreach \a in {90,162,234,306,378} {
        -- ++($(\a:-1.1*#1) + (\a+36:0.5*#1)$)
        -- ++($(\a+36:-0.5*#1) + (\a+72:1.1*#1)$)
      }
      -- cycle
    }
  },
  center arc/.style args={#1:#2:#3}{
    insert path={+ (#1:#3) arc (#1:#1+#2:#3)}
  }
}


\title{Nonuniform Buy-at-Bulk Network Design:\\ Improved Approximation and LP Integrality Gap}
\author{Iman Gholami\thanks{University of Maryland. Email: \texttt{igholami@umd.edu}.}
\and MohammadTaghi Hajiaghayi\thanks{University of Maryland. Email: \texttt{hajiagha@umd.edu}.}}
\date{}

\begin{document}
\hypersetup{pageanchor=false}
\begin{titlepage}
\maketitle
\thispagestyle{empty}
\begin{abstract}
In nonuniform buy-at-bulk network design, we are given an undirected graph and $h$ terminal pairs with positive demands, and we must route every demand at minimum total cost when each network resource has two independent parameters: a fixed buying price paid once if the resource is used, and a per-unit rental cost (or physical length) paid proportionally to the traffic it carries. Unlike the uniform-cost or single-sink variants, the general multicommodity problem allows unrelated buying and rental metrics across arbitrary source--sink pairs, arbitrary demands, and both edge and node resources.

We give a deterministic polynomial-time $O(\log^2 h / (\log\log h)^2)$-approximation algorithm for undirected nonuniform buy-at-bulk network design. The guarantee holds for edge costs, node costs, and their combination, and applies to arbitrary positive rational demands with no dependence on the number of vertices or the numerical demand ratio. This improves upon the $O(\log^4 h)$ bound of Chekuri, Hajiaghayi, Kortsarz, and Salavatipour (2010) for unrestricted instances. Our result also establishes the same $O(\log^2 h / (\log\log h)^2)$ upper bound on the integrality gap of the natural affine multicommodity flow LP, bounding integral buying and rental costs separately against their fractional counterparts and preserving zero rental budgets exactly. We complement this with an $\Omega(\log h/\log\log h)$ lower bound on the integrality gap of the natural LP that holds already for edge-only instances.
\end{abstract}

\end{titlepage}
\hypersetup{pageanchor=true}

\section{Introduction}
In many network design problems, building infrastructure and routing traffic over it incur two different kinds of cost. Installing a link or router requires a fixed upfront investment that is paid once regardless of how many connections share it, whereas routing traffic across an installed resource incurs a usage or delay cost that scales with the volume of traffic. Buy-at-bulk network design asks how to choose routes that balance these two forces: consolidating many demands onto a shared backbone saves installation cost, but routing a large demand along a long detour to reach that backbone can make its usage cost prohibitively high.

We study the undirected nonuniform multicommodity buy-at-bulk problem. The input consists of an undirected graph $G=(V,E)$ with $n=|V|$ vertices and $h$ source--destination pairs $(s_i,t_i)$ with positive rational demands $\delta_i>0$. Each network element (an edge, a vertex, or both) has two independent parameters: a fixed \emph{buying price} $b\ge0$, paid once if at least one path uses it, and a per-unit \emph{rental cost} (or physical length) $\ell\ge0$, paid by each demand that traverses it in proportion to its traffic $\delta_i$. A solution chooses an $s_i$--$t_i$ path $P_i$ for each request and pays
\begin{align*}
 &\underbrace{b\!\left(\bigcup_{i=1}^h P_i\right)}_{\text{buying paid once}}
 +\underbrace{\sum_{i=1}^h\delta_i\ell(P_i)}_{\text{rental paid by every demand}},
\end{align*}
where $b(\bigcup_{i=1}^h P_i)$ is the total buying price of the subgraph formed by the chosen paths and $\ell(P_i)$ is the physical length of path $P_i$. Crucially, buying an edge or vertex only pays its upfront installation price; its physical length never disappears, so every demand traversing an already bought element still pays its rental cost. We allow buying prices and physical lengths on edges, on nodes, or on both, with no common root or restriction on the placement of the terminal pairs, and no restriction on the numerical range of the demands.

Algorithmic progress on this general problem has advanced through several milestones. Charikar and Karagiozova~\cite{CK05} gave the first nontrivial guarantee, a randomized subpolynomial $\exp(O(\sqrt{\log n\log\log n}))$ approximation for unit demands on edges, with an additional logarithmic factor in the demand range for general demands. Chekuri, Hajiaghayi, Kortsarz, and Salavatipour~\cite{CHKS} then achieved the first polylogarithmic guarantee—an $O(\log^4 h)$ approximation for arbitrary demands and both node and edge costs—by clustering requests around shared intermediate roots in a junction-tree framework, and Kortsarz and Nutov~\cite{KN11} improved this junction-tree ratio to $O(\log^3 n)$ when demands are polynomially bounded.

Alongside these combinatorial and junction-tree algorithms, substantial effort has focused on bounding the integrality gap of the natural multicommodity LP relaxation. Because a low-cost junction tree in an integral routing does not by itself decompose a fractional multicommodity flow, the junction-tree bounds~\cite{CHKS,KN11} compare against the integral optimum rather than bounding the integrality gap of the natural multicommodity LP. Prior polylogarithmic upper bounds on the natural LP integrality gap were known in the single-sink setting—where $O(\log h)$ bounds hold for edge costs~\cite{MPP08,CKN01} and node costs~\cite{CHKS}—and, for multicommodity edge costs, through the recent work of Chekuri and Jain~\cite{CJ24}, who used hop-constrained oblivious routing~\cite{Raecke08,HHZ21,GHZ21,Filtser21} to round the natural edge-cost LP within a randomized factor of $O(\log D\,\log^3 n\,\log h)$, where $D=\max_i\delta_i$ in their demand convention. For the unrestricted multicommodity problem with arbitrary positive rational demands and node and edge costs, the $O(\log^4 h)$ approximation of~\cite{CHKS} has remained the best guarantee depending only on $h$, and no polylogarithmic bound depending only on $h$ was known for the natural LP relaxation.

On the lower-bound side, both edge-cost and node-cost instances exhibit logarithmic or polylogarithmic barriers. For the general concave edge-cost formulation, Andrews~\cite{Andrews04} ruled out $O(\log^{1/2-\varepsilon} n)$ approximations in the nonuniform case and $O(\log^{1/4-\varepsilon} n)$ approximations in the uniform case, for every fixed $\varepsilon>0$, unless $\mathrm{NP}\subseteq\mathrm{ZPTIME}(n^{\operatorname{polylog} n})$. Here $n$ counts vertices in that formulation; transferring these graph-size-dependent bounds through the affine reduction requires accounting for the size of the transformed instance. Chuzhoy, Gupta, Naor, and Sinha~\cite{CGNS08} showed $\Omega(\log\log n)$ hardness even for single-sink cost-distance, unless $\mathrm{NP}\subseteq\mathrm{DTIME}(n^{O(\log\log\log n)})$. For node costs, even when all physical lengths are zero, the problem captures node-weighted Steiner tree and set cover via the classical reduction of Klein and Ravi~\cite{KR95}; this yields both $(1-o(1))\ln h$ approximation hardness under $\mathrm{P}\ne\mathrm{NP}$~\cite{DS14} and an unconditional $\Omega(\log h)$ integrality-gap lower bound for the natural node-cost LP~\cite{Vondrak05}.

\subsection{Main results}\label{sec:main-results}
Our main result improves the approximation guarantee for the unrestricted problem from $O(\log^4 h)$ to $O(\log^2 h/(\log\log h)^2)$, with a factor that depends only on the number of terminal pairs $h$ and is independent of both the graph size $n$ and the numerical demand range.

\begin{theorem}[Unrestricted nonuniform buy-at-bulk]\label{thm:rounding}
There is a deterministic polynomial-time
\begin{align*}
 &O\!\left(\frac{\log^2 h}{(\log\log h)^2}\right)
\end{align*}
approximation algorithm for undirected nonuniform buy-at-bulk network design with $h\ge3$ arbitrary terminal pairs and positive rational demands. For one or two requests, the algorithm has a constant approximation ratio. The guarantee holds for buying prices and physical lengths on edges, on nodes, or on both, and establishes the same upper bound on the integrality gap of the natural affine multicommodity LP relaxation.
\end{theorem}

The approximation ratio in \Cref{thm:rounding} has no dependence on the number of vertices or the numerical ratio of the demands, and uses a single deterministic algorithm for edge costs, node costs, and their combination. By the standard edge-subdivision reduction~\cite{KR95,CHKS}, edge costs reduce directly to node costs on a transformed graph, which we again denote by $G=(V,E)$. Our algorithm rounds the natural affine LP relaxation, which opens each vertex $v\in V$ fractionally by $x_v\in[0,1]$ (or $x_e\in[0,1]$ for edges prior to subdivision) and sends one unit of $s_i$--$t_i$ flow $f_i$ for each request $i\in[h]$. If $f_i(v)$ denotes the total flow of request $i$ traversing vertex $v$, its sharing constraint is
\begin{align*}
 f_i(v)&\le x_v\qquad\text{for each request $i\in[h]$ and vertex $v\in V$,}
\end{align*}
and the objective minimizes the affine combination of fractional buying cost $B=\sum_{v\in V} b_v x_v$ and demand-weighted fractional rental cost $R=\sum_{i=1}^h \delta_i \sum_{v\in V} \ell_v f_i(v)$. These constraints allow the same fractional opening $x_v$ to support many different commodities, just as an integral solution pays the buying price $b_v$ once; they do not impose a shared capacity constraint $\sum_i f_i(v)\le x_v$.

Given any feasible fractional solution with buying cost $B$ and rental cost $R$, our rounding algorithm constructs a bought subgraph $F$ that separates the two budgets: for $h\ge3$,
\begin{align*}
 b(F)&\le O\!\left(\frac{\log^2 h}{(\log\log h)^2}\right)B,
 \qquad
 \sum_{i=1}^h\delta_i d_{F,\ell}(s_i,t_i)
 \le O\!\left(\frac{\log^2 h}{(\log\log h)^2}\right)R,
\end{align*}
where $d_{F,\ell}(s_i,t_i)$ is the shortest-path physical length between $s_i$ and $t_i$ in $F$ (and both bounds hold with $O(1)$ factors when $h\le2$). Because the two budgets are bounded separately, the rounding preserves a zero rental budget ($R=0$) or zero buying budget ($B=0$) exactly, rather than trading one cost against the other. \Cref{tab:comparison} summarizes how our guarantee compares with prior approximations for nonuniform multicommodity buy-at-bulk (where, as usual, prior-work bounds suppress constant lower cutoffs in logarithms).

\begin{table}[ht!]
\centering\small
\begin{tabular}{@{}>{\raggedright\arraybackslash}p{3.2cm}>{\raggedright\arraybackslash}p{3.4cm}>{\raggedright\arraybackslash}p{2.5cm}>{\raggedright\arraybackslash}p{4.1cm}@{}}
\toprule
Result & Resources and demands & Reference value & Approximation or gap upper bound\\
\midrule
Charikar--Karagiozova~\cite{CK05} & Edges; unit demands (log factor in demand range) & Integral optimum & $\exp(O(\sqrt{\log n\log\log n}))$, randomized\\
Chekuri et al.~\cite{CHKS} & Nodes and edges; arbitrary & Integral optimum & $O(\log^4 h)$\\
Kortsarz--Nutov~\cite{KN11} & Nodes and edges; polynomially bounded & Integral optimum & $O(\log^3 n)$\\
Chekuri--Jain~\cite{CJ24} & Edges; their demand convention & Natural LP & $O(\log D\,\log^3 n\,\log h)$, randomized\\
This work (\Cref{thm:rounding}) & Nodes and edges; arbitrary positive rationals & Natural LP and integral optimum & $O(\log^2 h/(\log\log h)^2)$, deterministic\\
\bottomrule
\end{tabular}
\caption{Comparison of approximations for nonuniform multicommodity buy-at-bulk network design. Here $n$ counts vertices, $h$ counts terminal pairs, and $D$ is the maximum demand in Chekuri and Jain's formulation.}
\label{tab:comparison}
\end{table}

We complement \Cref{thm:rounding} with a lower bound on the integrality gap of the natural affine LP that already holds when only edges carry buying prices and physical lengths. The classical $\Omega(\log h)$ set-cover gap (\Cref{sec:node-lower}) relies on node buying prices and does not yield such a gap for edge-only instances.

\begin{theorem}[Edge-only integrality gap]\label{thm:edge-gap}
There are infinitely many undirected edge-only instances of nonuniform buy-at-bulk network design, with $h\to\infty$ unit-demand terminal pairs and a graph of size polynomial in $h$, for which the natural affine LP relaxation has integrality gap
\begin{align*}
 &\Omega\!\left(\frac{\log h}{\log\log h}\right).
\end{align*}
In these instances, every edge has a strictly positive rational buying price and physical length with $O(\log h)$-bit encoding, every vertex has zero buying price and zero physical length, and the terminal pairs are pairwise vertex-disjoint.
\end{theorem}

Together with \Cref{thm:rounding}, \Cref{thm:edge-gap} shows that, for edge costs, the worst-case integrality gap of the natural LP is $\Omega(\log h/\log\log h)$ for infinitely many $h$ and $O(\log^2 h/(\log\log h)^2)$ for all $h\ge3$; the upper bound is thus at most the square of the lower bound, up to constant factors. \Cref{thm:edge-gap} concerns the LP relaxation only and is not a hardness-of-approximation result. Its proof in \Cref{sec:edge-lower} uses a high-girth regular demand graph, a polynomial-size covering lift with few cycles in small subgraphs, and a mod-$2$ dimension bound on requests routed along short connections.

\subsection{Related work}\label{sec:related}
The broader buy-at-bulk and network design literature studies several variations of the cost model, restricted demand patterns, and related algorithmic tools. These distinctions affect both the available approximation ratios and the lower bounds. While the theorems of this paper concern the two-parameter buying-and-rental model with unrestricted terminal pairs and demands, we survey these related models and techniques below to place our results in context.

\paragraph{Multiple cable types, uniform costs, and rent-or-buy.}
Buy-at-bulk network design is also studied with finitely encoded cable menus and with monotone concave or subadditive traffic-cost functions. Under the corresponding finite-input conventions, the constant-loss reduction of Chekuri, Hajiaghayi, Kortsarz, and Salavatipour~\cite[Section~2]{CHKS} approximates these functions by polynomially many affine buying-and-rental options; edge options are represented by parallel links, whereas node options require node copies. Our theorem concerns the resulting two-parameter affine model, and a general-tariff application additionally requires an appropriate finite representation of its cost functions. Salman, Cheriyan, Ravi, and Subramanian~\cite{SCRS97} initiated the study of single-sink edge installation with modular cable types. In the \emph{uniform} edge-cost variant, all edges share the same cost function up to an edge-dependent multiplier. Awerbuch and Azar~\cite{AA97} reduce uniform multicommodity buy-at-bulk to routing on probabilistically chosen trees; combined with the tree-metric embedding theorem of Fakcharoenphol, Rao, and Talwar~\cite{FRT04}, this gives an $O(\log n)$ approximation on $n$-vertex graphs. For edge costs, constant-factor approximations are known for uniform single-sink buy-at-bulk~\cite{GMM09} and multicommodity rent-or-buy~\cite{GKPR07}. In the latter model, an edge $e$ carrying load $x$ costs $c_e\min\{x,M\}$, where $M$ is a common threshold: one may rent capacity at rate $c_e$ per unit or buy unlimited capacity for $Mc_e$.

\paragraph{Nonuniform single-sink routing and multicast.}
Meyerson, Munagala, and Plotkin~\cite{MPP08} introduced the cost-distance formulation, which combines a network's installation cost with the distances from sources to a common sink. Their randomized logarithmic approximation allows the two edge metrics to be unrelated. Chekuri, Khanna, and Naor~\cite{CKN01} gave a deterministic LP-based algorithm, and Charikar, Naor, and Schieber~\cite{CNS04} studied closely related two-cost resource optimization in QoS multicast routing. Chekuri et al.~\cite{CHKS} extended the logarithmic approximation and the corresponding natural-LP bound to node costs. These results provide useful building blocks for junction-tree multicommodity algorithms~\cite{CHKS,KN11}, where groups of requests are routed through shared intermediate roots, but the common-sink assumption is essential to their single-sink guarantees. The rounding algorithm in this paper processes arbitrary pairs directly.

\paragraph{Oblivious routing with hop bounds.}
R\"acke~\cite{Raecke08} developed congestion-based tree decompositions for undirected routing. Hop constraints also require control of the number of edges on the paths that realize tree connections. Haeupler, Hershkowitz, and Zuzic~\cite{HHZ21}, Ghaffari, Haeupler, and Zuzic~\cite{GHZ21}, and Filtser~\cite{Filtser21} developed tree-embedding and oblivious-routing tools with such guarantees. Chekuri and Jain~\cite{CJ24} use the congestion-based version to round their network-design relaxation. Our proof uses ordinary shortest paths with nonnegative physical lengths. Its length analysis follows the actual paths used by each request, including the recursive arms that connect endpoints to centers.

\paragraph{Packing and sparse connection graphs.}
Berman and Coulston~\cite{BC97} use separated terminal centers to account for purchases in online Steiner forest. The auxiliary graphs here use the girth--density principle behind greedy sparse spanners~\cite{ADDJS93}; the nonregular Moore bound of Alon, Hoory, and Linial~\cite{AHL02} gives a stronger general graph estimate, and Alstrup, Dahlgaard, Filtser, St{\"o}ckel, and Wulff-Nilsen~\cite{ADFStW2019} construct light spanners deterministically in near-linear time. We prove the elementary girth estimate needed below. Center packing and high girth alone do not pay for the present algorithm: the attachment paths must also be purchased or realized through earlier routes. For node costs, disjoint geometric edge volume is unavailable. Our fixed toll rows and their deterministic cover provide the required replacement, and the proof charges center births directly to fractional buying.

\paragraph{Steiner problems, protected routing, and capacitated variations.}
Several classical and recent network design problems arise as specializations or close variants of nonuniform buy-at-bulk. When all physical lengths are zero ($\ell\equiv0$), only buying prices remain: for edge costs, the problem specializes to Steiner forest, which admits classic $2$-approximations via primal--dual methods~\cite{AKR95,GW95}, constant-factor greedy guarantees~\cite{GK15}, and recent $(2-\varepsilon)$-approximations~\cite{AGHJM25,GT26}; for node costs, it becomes node-weighted Steiner tree and Steiner forest, for which Klein and Ravi~\cite{KR95} gave an $O(\log h)$ approximation and subsequent work developed $O(\log n)$ approximations for prize-collecting and budgeted node-weighted variants~\cite{KSS13,BHL13}. In \emph{capacitated network design} and energy-efficient virtual circuit routing, edges carry hard capacities or speed-scaling costs rather than replicable buy-at-bulk cables; Chakrabarty, Chekuri, Khanna, and Korula~\cite{CCKK15} and Antoniadis et al.~\cite{AIKM20} study approximability and priority-based charging in those models. In \emph{buy-at-bulk network design with protection}, each demand pair requires fault-tolerant (edge- or node-disjoint) paths under buy-at-bulk economies of scale, studied by Antonakopoulos, Chekuri, Shepherd, and Zhang~\cite{ACSZ11} and recently shown to admit a polylogarithmic approximation by Chekuri and Jain~\cite{CJ26}. Finally, when stretch constraints replace additive rental costs in directed networks, Grigorescu, Kumar, and Lin~\cite{GKL25} study directed buy-at-bulk spanners.

\paragraph{Lower bounds and integrality gaps.}
The $\Omega(\log h)$ set-cover integrality-gap lower bound transferred via~\cite{KR95,Vondrak05} holds already on single-source instances with zero physical lengths (where the gap is tightly $\Theta(\log h)$), and together with \Cref{thm:rounding} places the worst-case integrality gap of the natural node-cost LP for general multicommodity instances between $\Omega(\log h)$ and $O(\log^2 h/(\log\log h)^2)$. By contrast, this node construction does not yield a logarithmic gap for the edge-only affine LP, where the computational hardness construction of Andrews~\cite{Andrews04} applies to general concave cost functions. For the edge-only LP, \Cref{thm:edge-gap} gives a separate $\Omega(\log h/\log\log h)$ integrality-gap lower bound (\Cref{sec:edge-lower}); no new hardness theorem is claimed here.

\subsection{Technical overview}
We process requests in nondecreasing order of their fractional physical lengths. For the current request, a mixed node-weighted shortest-path distance combines the buying price of resources not yet bought with their physical length. A common geometric grid of buying scales with ratio $\lambda$ determines the largest scale at which the original pair is still far apart under this distance.

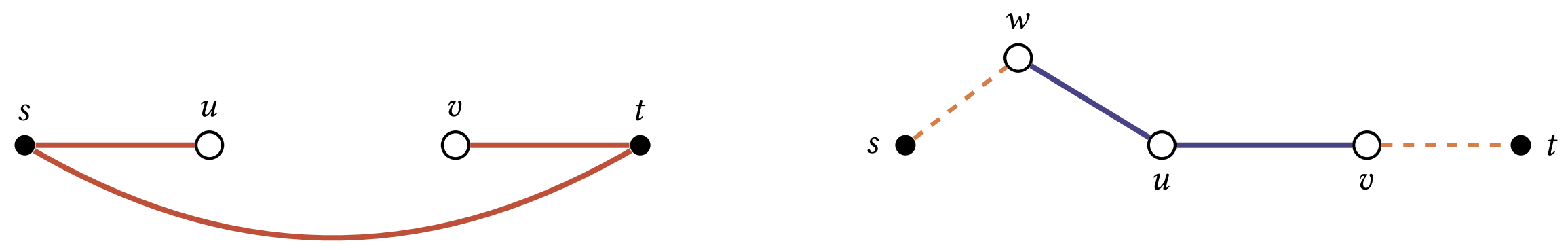
\begin{figure}[ht!]
\centering
\tikzset{
 overview terminal/.style={circle,fill=black,inner sep=2pt},
 overview center/.style={circle,draw=black,fill=white,thick,inner sep=2.6pt},
 overview bought/.style={Blue!75!Black,line width=1.5pt},
 overview purchase/.style={Red!75!Black,line width=1.5pt},
 overview pending/.style={Orange!85!Black,dashed,line width=1.2pt}
}
\begin{subfigure}[t]{0.48\textwidth}
\centering
\begin{tikzpicture}[font=\small]
\path[use as bounding box] (-.45,-1.05) rectangle (6.45,1.25);
\node[overview terminal,label=above:$s$] (s) at (0,0) {};
\node[overview center,label=above:$u$] (u) at (1.8,0) {};
\node[overview center,label=above:$v$] (v) at (4.2,0) {};
\node[overview terminal,label=above:$t$] (t) at (6,0) {};
\draw[overview purchase] (s)--(u);
\draw[overview purchase] (v)--(t);
\draw[overview purchase] (s) to[bend right=30] (t);
\end{tikzpicture}
\caption{Buy the main path and both arms.}
\label{fig:overview-a}
\end{subfigure}\hfill
\begin{subfigure}[t]{0.48\textwidth}
\centering
\begin{tikzpicture}[font=\small]
\path[use as bounding box] (-.45,-1.05) rectangle (6.45,1.25);
\node[overview terminal,label=left:$s$] (s) at (0,0) {};
\node[overview center,label=above:$w$] (w) at (1.1,.85) {};
\node[overview center,label=below:$u$] (u) at (2.5,0) {};
\node[overview center,label=below:$v$] (v) at (4.5,0) {};
\node[overview terminal,label=right:$t$] (t) at (6,0) {};
\draw[overview pending] (s)--(w);
\draw[overview bought] (w)--(u)--(v);
\draw[overview pending] (v)--(t);
\end{tikzpicture}
\caption{Reuse bought paths and attach the endpoints.}
\label{fig:overview-b}
\end{subfigure}
\caption{Two ways to connect a request $(s,t)$ through centers $u,v$. Red paths are bought now; blue paths are already bought. In (b), the left attachment uses another center $w$ and reuses its connection to $u$. Dashed arms are realized recursively. Each line represents a path in the network.}
\label{fig:overview}
\end{figure}

At that scale, each endpoint is assigned to a nearby existing center or becomes a center itself. If either center is newly recorded, or the two centers have no short bought connection, we buy a main path and the two attachment arms (\Cref{fig:overview-a}), recording one edge in an auxiliary graph for that scale. Otherwise, we reuse a short bought center connection and realize the attachment arms recursively at smaller scales (\Cref{fig:overview-b}), keeping the original endpoints as the starting points of these recursive attachments.

The buying proof has two parts. First, at each scale, separated center births consume disjoint tolls from fixed balls around original endpoints. For node costs, a deterministic cover of these toll rows is bought before processing requests. Its heavy and light node estimates bound the total birth count directly by the original fractional buying cost. Second, each recorded connection has a short physical realization. An edge inserted because of a center birth cannot close a cycle, since the newly eligible center has no earlier incident edge at that scale. For any other insertion, a cycle of length at most $2r$ would provide a short bought bypass, contradicting the insertion test. The resulting girth bound limits the number of recorded connections at each scale to $O((2h)^{1/r})$ times the number of births at that or larger scales. Summing their buying costs over the geometric grid gives the total buying bound.

The rental proof follows the actual physical route. Every skipped recursive level adds one bought connection of length at most $O(r)$ times the request's allowance, and every chain ends with a short attachment arm. With $K=\Theta(1+\log_\lambda h)$ active scales, the rental factor is $O(rK)$ while the main buying factor is $O(\lambda K(1+(2h)^{1/r}))$. A direct-completion test handles the case in which buying the first arm makes the original pair too close to justify another center birth; its extra buying cost is charged to buying already incurred during that request.

Finally, we space the common buying scales farther apart to trade the two factors against each other. Because the main path is computed at the next larger, ineligible scale, its physical length remains below the request's allowance regardless of the scale jump $\lambda$. Choosing the integer parameters
\begin{align*}
 q&=\min\{j\in\mathbb Z:j\ge2,\ j^j\ge2h\},\qquad
 r=2q,\qquad \lambda=\max\{8,\lceil\sqrt q\rceil\}
\end{align*}
gives $(2h)^{1/r}\le\sqrt{q}$ and $K=O(q)$ levels, so both the buying factor $O(\lambda K(2h)^{1/r})$ and the rental factor $O(rK)$ become $O(q^2)=O(\log^2 h/(\log\log h)^2)$.

\paragraph{Organization.}
\Cref{sec:prelim} defines the model and the natural LP. \Cref{sec:algorithm} gives the full algorithm and the recursive attachment invariant. \Cref{sec:helper} constructs the node-cost helper and proves the birth-packing bound. \Cref{sec:bounds} bounds buying and rental, including the high-girth argument. \Cref{sec:separate} handles zero budgets, and \Cref{sec:optimization} chooses the parameters to prove \Cref{thm:rounding}. \Cref{sec:edge-lower} proves the edge-only integrality-gap lower bound of \Cref{thm:edge-gap}. \Cref{sec:node-lower} verifies the classical node-LP integrality-gap lower bound for completeness.

\section{Model and relaxation}\label{sec:prelim}
We are given an undirected graph $G=(V,E)$ and $h$ terminal pairs $(s_i,t_i)$ for $i\in[h]:=\{1,\ldots,h\}$, each with a positive rational demand $\delta_i>0$. In the general model, each vertex $v\in V$ and edge $e\in E$ carries a nonnegative rational buying price ($b_v,b_e\ge0$) and a nonnegative rational physical length ($\ell_v,\ell_e\ge0$), with zero values on unweighted vertices or edges. All numerical input is encoded in binary. If an original pair is disconnected, the instance is infeasible; this can be detected before rounding. For a subgraph $F=(V(F),E(F))\subseteq G$ and a path $P$, write $b(F)=\sum_{v\in V(F)}b_v+\sum_{e\in E(F)}b_e$ and $\ell(P)=\sum_{v\in V(P)}\ell_v+\sum_{e\in E(P)}\ell_e$. An integral solution chooses a simple $s_i$--$t_i$ path $P_i$ for each request $i\in[h]$ and pays the total objective
\begin{align*}
 &b\!\left(\bigcup_{i=1}^h P_i\right)+\sum_{i=1}^h\delta_i\ell(P_i).
\end{align*}
For a bought subgraph $F\subseteq G$ and vertices $u,v$, we write $d_{F,\ell}(u,v)$ for the shortest-path physical length between $u$ and $v$ in $F$, and $d_{F,\ell}(u,v)=\infty$ if no $u$--$v$ path exists in $F$. Buying a vertex or edge removes its subsequent buying cost, but never its physical length. A bought graph may therefore support a long route whose rental is still expensive.

The case $h=0$ has optimum zero and is handled by returning the empty solution. Henceforth assume $h\ge1$ when defining logarithmic parameters and running the rounding algorithm.

Following the standard edge-subdivision reduction from edge costs to node costs in Klein and Ravi~\cite[Section~1]{KR95} and Chekuri, Hajiaghayi, Kortsarz, and Salavatipour~\cite[Section~2]{CHKS}, we reduce general instances with node and edge costs to node-cost instances ($b_e=\ell_e=0$ for all $e\in E$) with distinct terminal leaves.

\begin{lemma}\label{lem:node-reduction}
Any instance with node and edge costs can be reduced in polynomial time to an equivalent node-cost instance on $G=(V,E)$ with the same $h$ requests and distinct zero-cost, zero-length degree-one terminal leaves, preserving both the integral objective and its natural affine LP relaxation (\Cref{def:natural-lp}).
\end{lemma}
\begin{proof}
Subdivide each edge $e\in E$ with a degree-two vertex $v_e$ carrying $b_e$ and $\ell_e$~\cite[Section~1]{KR95}, \cite[Section~2]{CHKS}, and attach a distinct zero-cost, zero-length degree-one dummy leaf for each endpoint occurrence $s_i$ and $t_i$ ($i\in[h]$). All centers in our algorithm (\Cref{def:centers}) are chosen from these $2h$ terminal occurrences; because node lengths on a path include both endpoints, having zero physical length on each dummy leaf ensures that every center has self-distance zero. Map every original simple path to its subdivided path extended to the two endpoint leaves, and conversely suppress the subdivision vertices and strip the leaves. The path correspondence preserves the buying cost of every path union and the physical length of every path. Extend a feasible LP solution (with opening variables $x_v$ on vertices and $x_e$ on edges) by setting $x_{v_e}=x_e$ on each subdivision vertex and assigning opening one to the new zero-cost dummy leaves; conversely, project a transformed solution onto the original vertices, edges, and corresponding paths. These maps preserve feasibility and both budgets $(B,R)$, so both the integral and fractional optimum values are preserved.
\end{proof}

By \Cref{lem:node-reduction}, for the remainder of the paper all buying prices and physical lengths reside on vertices $v\in V$, so $b(S)=\sum_{v\in S}b_v$ for $S\subseteq V$ and $\ell(P)=\sum_{v\in P}\ell_v$ for a path $P$.

\begin{definition}[Natural affine LP]\label{def:natural-lp}
For each request $i\in[h]$, let $\mathcal P_i$ denote the set of simple $s_i$--$t_i$ paths in $G$. The \emph{natural affine LP} relaxation of the buy-at-bulk problem (see~\cite{CHKS,CJ24} for related formulations) associates a fractional opening variable $x_v\in[0,1]$ with each vertex $v\in V$ and a path-flow variable $f_{iP}\ge0$ with each request $i\in[h]$ and path $P\in\mathcal P_i$, and solves
\begin{align*}
 \min\quad &\sum_{v\in V} b_v x_v + \sum_{i=1}^h\delta_i\sum_{P\in\mathcal P_i}\ell(P)f_{iP}\\
 \text{s.t.}\quad
 &\sum_{P\in\mathcal P_i}f_{iP}=1 &&(i\in[h]),\\
 &f_i(v):=\sum_{P\in\mathcal P_i:v\in P}f_{iP}\le x_v &&(i\in[h],\ v\in V),\\
 &0\le x_v\le1,\qquad f_{iP}\ge0 &&(i\in[h],\ v\in V,\ P\in\mathcal P_i).
\end{align*}
For a feasible fractional solution $(x,f)$, we write $\ell(f_i):=\sum_{P\in\mathcal P_i}\ell(P)f_{iP}=\sum_{v\in V}\ell_v f_i(v)$ for the fractional physical length of request $i$, and define the \emph{fractional buying cost} $B$, \emph{per-request rental cost} $R_i$, and \emph{fractional rental cost} $R$ by
\begin{align*}
 B&:=\sum_{v\in V}b_vx_v,\qquad R_i:=\delta_i\ell(f_i),\qquad R:=\sum_{i=1}^h R_i.
\end{align*}
\end{definition}

Restricting $x_v,f_{iP}\in\{0,1\}$ in \Cref{def:natural-lp} recovers the exact integral problem. In the linear relaxation, each vertex $v\in V$ incurs the affine cost $b_v x_v + \ell_v\sum_{i=1}^h \delta_i f_i(v)$ subject to a separate capacity-sharing constraint $f_i(v)\le x_v$ for each request $i\in[h]$; there is no aggregate capacity constraint on $\sum_{i=1}^h f_i(v)$. The next lemma gives an explicit compact formulation. Standard node splitting~\cite[Section~2.4]{AMO93} and the flow-decomposition theorem support its construction and path representation.

\begin{lemma}[Polynomial flow formulation]\label{lem:lp-solution}
The natural affine LP (\Cref{def:natural-lp}) can be solved in polynomial time. Given a feasible rational solution represented by a compact commodity flow or an explicit path list, a representation on polynomially many simple paths can be computed without increasing its buying or rental cost, in time polynomial in the input representation's encoding length.
\end{lemma}
\begin{proof}
After the endpoint-leaf reduction, $s_i\ne t_i$ for every request. Replace every undirected edge $uv$ by directed transit arcs $uv$ and $vu$. For each request, let $g_i(u,v)\ge0$ be its transit flow and let $y_i(v)$ be its node throughput. Use the constraints
\begin{align*}
 y_i(v)&=\sum_{u:uv\in E}g_i(u,v)+\mathbf 1\{v=s_i\},\\
 y_i(v)&=\sum_{u:uv\in E}g_i(v,u)+\mathbf 1\{v=t_i\},\\
 0&\le y_i(v)\le x_v\le1
 \qquad(i\in[h],\ v\in V),
\end{align*}
and objective $\sum_v b_vx_v+\sum_i\delta_i\sum_v\ell_v y_i(v)$. This LP has $O(|V|+h(|V|+|E|))$ variables and constraints and rational coefficients of polynomial encoding length, so rational linear programming solves it in polynomial time.

Every feasible path solution induces these transit flows and throughputs with exactly the same objective. Conversely, conservation gives a unit directed $s_i$--$t_i$ flow for each request. By the flow-decomposition theorem~\cite[Theorem~3.5]{AMO93}, it decomposes into simple paths and cycles. Delete all cycles; nonnegative costs ensure that throughput and rental only decrease. The remaining path flows sum to one, use each node to extent at most $y_i(v)\le x_v$, and hence satisfy \Cref{def:natural-lp}. There are polynomially many path and cycle components, obtained by repeatedly extracting a positive-flow path or cycle and zeroing at least one arc. Thus the compact and path formulations have the same optimum, and the desired path representation is available in polynomial time. An explicit supplied path list can first be summed into compact transit flows and treated in the same way. For each commodity, let $D$ be the product of the denominators of its input arc flows. Its binary length is bounded by the total input denominator length. Minimum selection and residual subtraction preserve multiples of $1/D$, and all residual arc flows remain between zero and one. Hence the extracted weights and residual flows have polynomial encoding length. The initial aggregation of an explicit path list is polynomial in that list's encoding length.
\end{proof}

\section{The rounding algorithm}
\label{sec:algorithm}

Fix integers $\lambda\ge8$ and $r\ge1$, and set
\begin{align*}
 K&=\left\lceil\log_\lambda(2\lambda h)\right\rceil,
 \qquad J=\lceil\log_2(2h)\rceil.
\end{align*}
We now describe the core parameterized rounding run for a subset of requests equipped with prescribed physical length allowances $L_i>0$ satisfying $\ell(f_i)\le L_i/64$, assuming $B>0$. (\Cref{sec:separate} handles zero buying budgets and zero-rental requests first, and then invokes this run on the remaining requests with $L_i=64R_i/\delta_i=64\ell(f_i)$ to establish the parameterized guarantee in \Cref{thm:main}; \Cref{sec:optimization} then chooses $\lambda$ and $r$ to obtain \Cref{thm:lp-rounding,thm:rounding}.)
An arbitrary previously bought seed graph is permitted. Add a helper set $S$ of buying cost $O(\log(2h))B$, constructed in \Cref{sec:helper}, and all zero-buying
resources. Initialize fresh center records and fresh auxiliary
graphs. Let $Q=6r+6$. Only newly bought resources are charged to this run. Each request is processed in nondecreasing
$L_i$; ties use any fixed order.

\begin{definition}[Center records and eligibility]\label{def:centers}
A \emph{center record} consists of an original endpoint and its birth scale; creating a new record at scale $\tau$ is called a \emph{center birth at scale $\tau$}. A physical endpoint is \emph{eligible as a center at scale $\tau$} if it has a record at a scale at least $\tau$. Recording an endpoint at a larger scale than its existing records is a \emph{promotion} (and is counted among the births at that larger scale). Eligibility persists as resources are bought.
\end{definition}

\begin{definition}[Mixed shortest-path distance and pair eligibility]\label{def:mixed-metric}
For the current request $i$, the \emph{residual buying price} $b_v^{\mathrm{res}}$ is zero on bought nodes and $b_v$ otherwise. At scale $\tau$, the mixed node cost is
\begin{align*}
 w_\tau(v)&=b_v^{\mathrm{res}}+(\tau/L_i)\ell_v.
\end{align*}
Write $d_{w_\tau}$ for the node-weighted shortest-path distance in these node costs (summing over all vertices on a path, including both endpoints; since all terminal leaves and centers have zero cost and length by \Cref{lem:node-reduction}, every center has self-distance zero). A scale is \emph{eligible for the original pair} at the start of its request if $d_{w_\tau}(s_i,t_i)\ge\tau$ at that time. This pair-scale test is distinct from center eligibility in \Cref{def:centers}.
\end{definition}

\subsection{Scale choice and the original-pair witness}

For the current request abbreviate $L_i$ by $L$. The following lemma justifies searching the grid for the largest eligible pair scale.

\begin{lemma}[Original-pair scale witness]\label{lem:scale-witness}
At the start of a request, the top scale $2B$ is ineligible. If a scale $\tau_*$ is eligible, every positive scale at most $\tau_*$ is eligible at that time.
\end{lemma}
\begin{proof}
The LP comparator gives
\begin{align*}
 d_{w_\tau}(s,t)
 &\le\sum_P f_{iP}w_\tau(P)
 \le B+\tau/64.
\end{align*}
The buying term is at most $B$ because $f_i(v)\le x_v$ for every node; the physical term uses $\ell(f_i)\le L_i/64$. At $\tau=2B$ the right side is strictly less than $2B$. For a fixed path,
$w_\tau(P)/\tau=b^{\mathrm{res}}(P)/\tau+\ell(P)/L_i$ is nonincreasing in $\tau$, and its minimum over paths has the same property, so $d^{\mathrm{start}}_{w_\tau}(s,t)\ge\tau$ for all $0<\tau\le\tau_*$.
\end{proof}

\begin{lemma}[Distance stability within one request]
\label{lem:stability}
If actual new buying during the current request has totaled
$C_i^{\mathrm{pre}}$, then
\begin{align*}
 d^{\mathrm{current}}_{w_\tau}(s,t)
 &\ge d^{\mathrm{start}}_{w_\tau}(s,t)-C_i^{\mathrm{pre}}.
\end{align*}
\end{lemma}
\begin{proof}
Take a simple current shortest path. Restoring the buying prices
removed during this request adds at most $C_i^{\mathrm{pre}}$,
because the path uses each node at most once. Its physical length
does not change. Comparing with the old shortest distance proves the lemma. Purchases made before the request, including the helper, are
absent from this expenditure counter.
\end{proof}

\subsection{Main step}

\begin{figure}[t]
\centering
\def\dem{Red!70}
\def\demi{Blue!70}
\def\col{Black!70}
\begin{subfigure}[b]{0.53\textwidth}
\centering
\begin{tikzpicture}[>=Stealth,scale=.95,
  term/.style={circle,draw=\col,fill=\col,line width=1.0pt,inner sep=1.8pt},
  ctr/.style={circle,draw=\col,fill=white,line width=1.0pt,inner sep=2.2pt}]
\path (0,-1.45) rectangle (0,0.65);
\node[term,label={[text=\col]below:$s$}] (s) at (0,0) {};
\node[ctr,label={[text=\col]below:$u$}] (u) at (2.4,0) {};
\node[ctr,label={[text=\col]below:$v$}] (v) at (5.4,0) {};
\node[term,label={[text=\col]below:$t$}] (t) at (7.8,0) {};
\draw[->,\dem,line width=1.4pt] (s)--node[above,font=\small,text=\col] {$P_s$} (u);
\draw[\demi,line width=1.6pt] (u)--node[above,font=\small,text=\col] {$W$} (v);
\draw[<-,\dem,line width=1.4pt] (v)--node[above,font=\small,text=\col] {$P_t$} (t);
\draw[\col,dashed,line width=1.1pt] (s) to[bend right=18] node[below,font=\small,text=\col] {$P_0$} (t);
\end{tikzpicture}
\caption{Main step at scale $\tau_*$.}
\label{fig:routes-a}
\end{subfigure}\hfill
\begin{subfigure}[b]{0.43\textwidth}
\centering
\begin{tikzpicture}[>=Stealth,scale=.95,
  term/.style={circle,draw=\col,fill=\col,line width=1.0pt,inner sep=1.8pt},
  ctr/.style={circle,draw=\col,fill=white,line width=1.0pt,inner sep=2.2pt}]
\path (0,-1.45) rectangle (0,0.65);
\node[term,label={[text=\col]below:$s$}] (s2) at (0,0) {};
\node[ctr,label={[text=\col]below:$v'$}] (vp) at (2.5,0) {};
\node[ctr,label={[text=\col]below:$u$}] (u2) at (5.5,0) {};
\draw[->,\dem,line width=1.4pt] (s2)--node[above,font=\small,text=\col] {$P'$} (vp);
\draw[\demi,line width=1.6pt] (vp)--node[above,font=\small,text=\col] {$W'$} (u2);
\draw[\col,dashed,line width=1.1pt] (s2) to[bend right=25] node[below,font=\small,text=\col] {$P_s$} (u2);
\end{tikzpicture}
\caption{Recursive step at scale $\tau=\tau_*/\lambda$.}
\label{fig:routes-b}
\end{subfigure}
\caption{Choices in \textsc{ServeDemandPair} and \textsc{ConnectEndpoint}. (\subref{fig:routes-a}) At the largest eligible scale $\tau_*$, endpoints $s,t$ attach to centers $u,v$: the algorithm either buys $P_s\cup P_0\cup P_t$ or reuses a short bought route $W$ and attaches the arms recursively. (\subref{fig:routes-b}) At scale $\tau=\tau_*/\lambda$, the recursive step from the original endpoint $s$ finds a closer center $v'$ and either buys $P_s\cup P'$ or appends a bought route $W'$ and recurses on $(s,v')$.}
\label{fig:routes}
\end{figure}
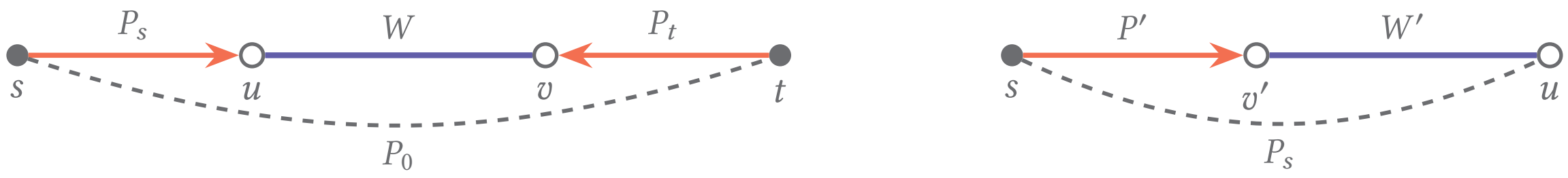

The procedure \textsc{ServeDemandPair} (\Cref{alg:main}) either buys a short path or reuses a short connection between centers (\Cref{fig:routes-a}). It assigns both endpoints before buying, so the original pair witnesses both initial assignments. Set $\tau_k=2B\lambda^k$ for $-K\le k\le0$ and $\eta=B/(\lambda h)$.

\begin{algorithm}[t]
\small
\KwIn{Request $(s,t)$, allowance $L=L_i$, bought graph $F$, center records, auxiliary graphs.}
\KwOut{A bought route for this request; updated $F$, records, and auxiliary graphs.}
\caption{ServeDemandPair: connect a request by buying or reusing center paths}\label{alg:main}
\If{$d_{F,\ell}(s,t)\le L$}{Serve the request on a shortest bought path; \KwRet\;}
Inspect the grid and find its eligible scales in the current mixed distance\;
\If{no scale is eligible}{Buy a shortest mixed path at $\tau_{-K}$; serve the request on it; \KwRet\;}
Let $\tau_*$ be the largest eligible scale\;
Compute a shortest mixed path $P_0$ at $\lambda\tau_*$\tcp*[r]{Use the larger ineligible scale.}
\If{$\tau_*\le\eta$}{Buy $P_0$; serve the request on it; \KwRet\;}
\tcp{Assign both endpoints before buying so the original pair witnesses both searches.}
For each endpoint, find an eligible center at mixed distance at most $\tau_*/8$\;
If none exists, record the endpoint itself as a center at $\tau_*$\;
Let $u,v$ be the assigned centers and $P_s,P_t$ their shortest arms\tcp*[r]{No buying yet.}
\If{a center is new, or $d_{F,\ell}(u,v)>QL$}{
 Buy $P_s\cup P_0\cup P_t$ and record $uv$ at scale $\tau_*$\;
 Store its physical realization through these paths\tcp*[r]{Keep lengths for later bypass tests.}
 Serve the request on $P_0$; \KwRet\;
}
Retain a bought $u$--$v$ path $W$ of physical length at most $QL$\;
Call $\operatorname{ConnectEndpoint}(s,u,P_s,\tau_*/\lambda)$\;
\If{the call reports completion}{\KwRet\;}
Call $\operatorname{ConnectEndpoint}(t,v,P_t,\tau_*/\lambda)$\;
\If{the call reports completion}{\KwRet\;}
Serve the request on the two returned arms joined by $W$, erasing cycles\;
\end{algorithm}

\begin{lemma}[Main-step paths]\label{lem:main-paths}
The larger scale used by \Cref{alg:main} exists in the grid. Its main path satisfies
\begin{align*}
 w_{\lambda\tau_*}(P_0)&<\lambda\tau_*,\qquad
 b^{\mathrm{res}}(P_0)<\lambda\tau_*,\qquad \ell(P_0)<L.
\end{align*}
The assigned centers are distinct, and both attachment arms have mixed cost at most $\tau_*/8$ and physical length at most $L/8$. Every recorded main edge costs at most $(\lambda+1/4)\tau_*$ in new buying and has a physical realization shorter than $3L$. The cheap main branch costs at most $B/h$.
\end{lemma}
\begin{proof}
By \Cref{lem:scale-witness}, the top scale is ineligible. The grid scale immediately above the largest eligible scale is therefore present and ineligible. Its shortest path gives the stated bounds on $P_0$, using nonnegativity of buying and physical lengths. An arm of mixed cost at most $\tau_*/8$ has physical length at most $L/8$. A newborn center has a zero arm. If the centers coincided, concatenating the arms would contradict eligibility of the original pair. Buying both arms and $P_0$ costs at most $(\lambda+1/4)\tau_*$; their concatenation realizes the auxiliary edge with length less than $5L/4<3L$. If no scale is eligible, the smallest-scale path costs less than $\tau_{-K}\le\eta$ and has physical length less than $L$. If $\tau_*\le\eta$, the bound $b^{\mathrm{res}}(P_0)<\lambda\tau_*$ bounds buying by $\lambda\eta=B/h$.
\end{proof}

\subsection{Attachment subroutine and its invariant}

The procedure \textsc{ConnectEndpoint} keeps the original endpoint as its starting point while replacing its target center at smaller scales (\Cref{fig:routes-b}).

\begin{definition}[Inherited endpoint arm]\label{def:inherited-arm}
An \emph{inherited arm} at scale $\tau$ is a physical path $P$ from an original endpoint $z$ to a center $u$ eligible at $\tau$, satisfying
\begin{align*}
 w_\tau(P)&\le(\lambda/8)\tau,\qquad \ell(P)\le L/8.
\end{align*}
for the current request allowance $L$.
\end{definition}

The completion signal below ends the whole original request; pending attachment calls do not resume after that signal.

\begin{algorithm}[ht!]
\small
\KwIn{Original endpoint $z$, center $u$ eligible at $\tau$, inherited arm $P$ satisfying \Cref{def:inherited-arm}; current request $(s,t)$ and allowance $L$.}
\KwOut{A bought $z$--$u$ route, or a signal that the original request has been completed.}
\caption{ConnectEndpoint: realize an endpoint arm through bought connections}\label{alg:attach}
\If{$\tau\le\eta$}{Buy $P$; \KwRet $P$\;}
\If{$d_{w_\tau}(s,t)\le\tau/16$}{\label{line:complete}
 Buy a shortest mixed $s$--$t$ path and serve the original request on it\;
 \KwRet completion\tcp*[r]{Test the original pair, not two centers.}
}
\If{no eligible center lies within mixed distance $\tau/8$ of $z$}{\label{line:birth}
 \tcp{A separated endpoint creates a record paid for by the packing bound.}
 Record $z$ as a center at $\tau$; buy $P$\;
 Record auxiliary edge $zu$ at $\tau$, realized by $P$; \KwRet $P$\;
}
Choose a nearby center $v$ and a shortest mixed $z$--$v$ arm $P'$\;
\If{$d_{F,\ell}(u,v)>QL$}{\label{line:purchase}
 \tcp{Both arms are bought so the recorded center edge has a physical route.}
 Buy $P\cup P'$; record $uv$ at $\tau$, realized through $P,P'$\;
 \KwRet $P$\;
}
Retain a bought $v$--$u$ route $W$ of physical length at most $QL$\;
Call $\operatorname{ConnectEndpoint}(z,v,P',\tau/\lambda)$\tcp*[r]{Lower buying scale; keep the same endpoint.}
\If{the call reports completion}{\KwRet completion\;}
\KwRet its returned route followed by $W$, with cycles erased\;
\end{algorithm}

\begin{lemma}[Attachment invariant and event bounds]\label{lem:attachment}
The initial and recursive calls satisfy \Cref{def:inherited-arm}. A completion costs at most $C_i^{\mathrm{pre}}/15$, where $C_i^{\mathrm{pre}}$ is the buying already incurred during this request, and has physical length at most $L/16$. Every actual birth has its original mate at mixed distance greater than $\tau/16$. Every non-base, non-completion purchase records one auxiliary edge, with new buying and realization length at most
\begin{align*}
 &(\lambda+1/4)\tau\quad\text{and}\quad 3L,
\end{align*}
respectively. Each request creates at most two center records. The total buying of all cheap main events and attachment bases is at most $B$.
\end{lemma}
\begin{proof}
An initial main arm has mixed cost at most $\tau_*/8$. Lowering its scale to $\tau_*/\lambda$ cannot increase that cost and gives \Cref{def:inherited-arm}. For a recursive arm,
\begin{align*}
 w_{\tau/\lambda}(P')&\le w_\tau(P')\le\tau/8
                 =(\lambda/8)(\tau/\lambda),
\end{align*}
while its physical length is at most $L/8$. This proves preservation.

All attachment scales lie below the initially eligible scale. By \Cref{lem:scale-witness} and \Cref{lem:stability}, a completion test that succeeds implies $C_i^{\mathrm{pre}}\ge15\tau/16$. Its new buying is therefore at most
\begin{align*}
 \tau/16&\le C_i^{\mathrm{pre}}/15,
\end{align*}
and its physical length is at most $L/16$. The completion signal ends the request, so this event occurs at most once. An attachment birth occurs only after the original-pair test failed; main births occur before any within-request purchase and have the initial eligibility witness.

The birth branch buys an inherited arm for at most $\lambda\tau/8$ and realizes its edge with length at most $L/8$. The other purchase branch buys that arm and a near arm, costing at most $(\lambda+1)\tau/8$ with realization length at most $L/4$. Together with \Cref{lem:main-paths}, these estimates prove the stated per-edge buying and length bounds.

A skipped level makes neither a purchase nor a birth. An attachment chain has just one terminal purchase and can create a center only at that terminal event. There are at most two chains per request; a main birth instead ends the request without attachment calls. Hence at most two center records are created per request. A base costs at most $(\lambda/8)\eta=B/(8h)$. Each request has either one cheap main event costing at most $B/h$ or at most two such bases. Summing over at most $h$ requests bounds these cheap purchases by $B$.
\end{proof}

\section{A deterministic fixed-ball cover for node costs}
\label{sec:helper}

We now construct the helper set and use it to bound the number of centers created at each scale.
Assume $B>0$. Consider any subset of the original requests with
positive allowances $L_i$ satisfying
\begin{align*}
 \ell(f_i)&\le L_i/64.
\end{align*}
Use the common scales
\begin{align*}
 \tau_k&=2B\lambda^k\quad(-K\le k\le0),\qquad
 \eta=B/(\lambda h).
\end{align*}
\begin{lemma}[Finite grid]\label{lem:grid}
The grid has at most $K$ scales strictly above $\eta$, its smallest scale satisfies $\tau_{-K}\le\eta$, and it produces at most $(8h)^2$ endpoint-scale rows.
\end{lemma}
\begin{proof}
The definition of $K$ gives $\lambda^K\ge2\lambda h$, hence $\tau_{-K}\le B/(\lambda h)=\eta$. Moreover,
\begin{align*}
 K&=1+\lceil\log_\lambda(2h)\rceil\le J+1.
\end{align*}
There are at most $2h$ endpoint occurrences and $K+1$ grid scales. Thus their product is at most $2h(J+2)\le(8h)^2$ for $h\ge1$.
\end{proof}

\subsection{Original toll rows and their fractional mass}

\begin{definition}[Fixed toll rows]\label{def:toll-rows}
For an endpoint $s$ of request $i$ and a grid scale $\tau$, set
$\rho=\tau/16$. Let $d^0_{i,\tau}(s,v)$ be the node-weighted shortest-path distance in the
\emph{original} node costs
$b_v+(\tau/L_i)\ell_v$, before any purchases. Include the row
$a=(i,s,\tau)$ only if its mate $t$ satisfies
$d^0_{i,\tau}(s,t)>\rho$. Its truncated buying toll is
\begin{align*}
 z_a(v)&=\max\{0,\min\{b_v,\rho-d^0_{i,\tau}(s,v)+b_v\}\}.
\end{align*}
Set the toll to zero at unreachable nodes. In particular,
$0\le z_a(v)\le\min\{b_v,\rho\}$.

\end{definition}

\begin{lemma}[Fractional toll mass]\label{lem:toll-mass}
Every included row satisfies
\begin{align*}
 \sum_vx_vz_a(v)&\ge3\rho/4.
\end{align*}
\end{lemma}
\begin{proof}
Traverse any positive original LP path $P=(s=v_0,v_1,\ldots,v_m=t)$ from $s$ to its mate $t$, and abbreviate $d(s,v)=d^0_{i,\tau}(s,v)$. By the shortest-path prefix inequality, every consecutive pair $(u,v)$ along $P$ satisfies $d(s,v)\le d(s,u)+b_v+(\tau/L_i)\ell_v$. Consequently, the truncated distance $\min\{d(s,\cdot),\rho\}$ satisfies
\begin{align*}
 \min\{d(s,v),\rho\}-\min\{d(s,u),\rho\}
 &\le z_a(v)+(\tau/L_i)\ell_v.
\end{align*}
Indeed, if $d(s,u)\ge\rho$ the left-hand side is at most zero while the right-hand side is nonnegative; if $d(s,v)\le\rho$ then $z_a(v)=b_v$; and if $d(s,u)<\rho<d(s,v)$ then $\rho-d(s,u)\le(\rho-d(s,v)+b_v)+(\tau/L_i)\ell_v\le z_a(v)+(\tau/L_i)\ell_v$ by \Cref{def:toll-rows}. Since the dummy terminal leaf $s$ has $b_s=\ell_s=0$ (\Cref{lem:node-reduction}) and the included row satisfies $d(s,t)>\rho$, the truncated distance starts at $\min\{d(s,s),\rho\}=0$ and ends at $\min\{d(s,t),\rho\}=\rho$. Summing these inequalities over the steps of $P$ and telescoping gives
\begin{align*}
 \rho&\le\sum_{v\in P}z_a(v)+(\tau/L_i)\ell(P).
\end{align*}
Average over the unit flow $f_i$, use $f_i(v)\le x_v$, and then
use the allowance condition $\ell(f_i)\le L_i/64$. The subtracted physical contribution is at most
$\tau/64=\rho/4$.
\end{proof}

\subsection{Heavy and light nodes}

\begin{definition}[Heavy nodes and toll rows]\label{def:heavy-light}
Set
\begin{align*}
 \alpha&=128\lceil\log_2(8h)\rceil,
 \qquad H=\{v:\alpha x_v\ge1\},\qquad V_{\mathrm{light}}=V\setminus H.
\end{align*}
A row is heavy if
$\sum_{v\in H}x_vz_a(v)\ge3\rho/8$; otherwise it is light.
\end{definition}

Initially sample each light node independently with probability
$p_v=\alpha x_v$.

\begin{definition}[Pessimistic estimator]\label{def:estimator}
The procedure \textsc{CoverFixedTolls} chooses light nodes by fixing independent selection indicators one at a time. Let $X_v$ indicate selection of light node $v$, and let $S$ consist of the heavy nodes and the selected light nodes. Define
\begin{align*}
 \psi_a&=2^{\alpha/8}
        \prod_{v\in V_{\mathrm{light}}}\left(1-\frac{z_a(v)}{2\rho}\right)^{X_v},
 \qquad
 \Phi=\frac{b(S)}{4\alpha B}+\sum_{a\text{ light}}\psi_a.
\end{align*}
\end{definition}

The first term pays for buying; a light row with too little selected toll will force its corresponding $\psi_a$ to be large. The algorithm fixes a selection without increasing the conditional expectation of $\Phi$.

\begin{algorithm}[ht!]
\small
\KwIn{Original toll rows $z_a$, openings $x_v$, buying prices $b_v$, and $B>0$.}
\KwOut{The helper set $S$ guaranteed by \Cref{lem:helper}.}
\caption{CoverFixedTolls: buy nodes that cover every light toll row}\label{alg:helper}
Compute $\alpha$, the heavy node set $H$, and the light rows\;
Set $S\gets H$; for light nodes use independent indicators with probabilities $p_v=\alpha x_v$\;
Fix indicators of probability zero to zero\;
\ForEach{remaining unfixed light-node indicator $X_v$}{
 Compute $A_0=\E[\Phi\mid\text{previous choices},X_v=0]$ and $A_1=\E[\Phi\mid\text{previous choices},X_v=1]$\;
 Fix $X_v$ to a value minimizing $A_0,A_1$\tcp*[r]{Keep the conditional expectation from increasing.}
}
Add all light nodes with $X_v=1$ to $S$; buy $S$\;
\KwRet $S$\;
\end{algorithm}

\begin{lemma}[Exact deterministic helper]
\label{lem:helper}
There is a deterministic polynomial-time choice of a set $S$
containing all heavy nodes such that
\begin{align*}
 b(S)&<4\alpha B,
 \qquad
 \sum_{v\in S\cap V_{\mathrm{light}}}z_a(v)>\alpha\rho/8
       \quad\hbox{for every light row }a.
\end{align*}
\end{lemma}
\begin{proof}
For a light row, \Cref{lem:toll-mass} and the heavy-row threshold imply $\sum_{v\in V_{\mathrm{light}}}x_vz_a(v)>3\rho/8$. 
For $0\le t\le1$, concavity of $\log(1-t/2)$ gives
$1-t/2\ge2^{-t}$. Therefore, if the selected toll in a light row
is at most $\alpha\rho/8$, then $\psi_a\ge1$.
On the other hand, independence and $1-u\le e^{-u}$ give
\begin{align*}
 \E\psi_a
 &\le\exp\!\left(\frac{\alpha\ln2}{8}
            -\frac{\alpha}{2\rho}\sum_{v\in V_{\mathrm{light}}}x_vz_a(v)\right)
 \le e^{-\alpha/16}.
\end{align*}
The expected buying cost, including heavy nodes, is at most
$\alpha B$. Consequently the estimator $\Phi$ has expectation below $1/2$: there are at most $(8h)^2$ rows,
and the chosen value of $\alpha$ makes their contribution below
$1/4$.

Fix probability-zero and probability-one variables. Sequentially
fix each remaining variable to a branch of no greater conditional
expectation. This uses only rational arithmetic: $\alpha/8$ is an
integer, the factors in \Cref{def:estimator} lie in $[1/2,1]$, and each conditional
expectation is an explicit finite product of rational factors.
The final $\Phi$ remains below $1/2$. Its nonnegative cost term
implies $b(S)<4\alpha B$; a failed light row would
contribute at least one, proving the toll-coverage bound. The number and bit
length of all factors are polynomial in the rational input size.
\end{proof}

\subsection{Bounding center births}

A birth at endpoint $s$ and scale $\tau$ will be allowed only if
there is no eligible center at mixed distance at most $\tau/8$
from $s$, and the current distance from $s$ to its original mate
is greater than $\tau/16$.

\begin{lemma}[Constant-coefficient node packing]
\label{lem:packing}
Let $N_k$ count actual birth records at scale $\tau_k$, including
any promotions. Under the preceding rules,
\begin{align*}
 \tau_kN_k&\le C_0B,\qquad C_0=1664/3.
\end{align*}
\end{lemma}
\begin{proof}
Every actual birth has an included original row: its original mate
is outside the current radius $\tau_k/16$, and original buying
prices only make that distance larger.

At a fixed scale the positive-toll supports on selected nodes are
disjoint among actual birth rows (\Cref{fig:buying-a}). Suppose a selected node $v$ had
positive toll for an earlier and a later birth. For each original
row, \Cref{def:toll-rows} supplies an arm to $v$ whose original mixed cost after
excluding $b_v$ is strictly less than $\rho=\tau_k/16$.
Node $v$ is already bought. All other buying prices only decrease,
and the later allowance is at least the earlier one. Thus in the
later mixed distance these arms concatenate to a route of cost
strictly below $2\rho=\tau_k/8$ from the new endpoint to the old
center. The old center remains eligible. This contradicts the new
birth rule.

For heavy birth rows, disjointness, \Cref{lem:toll-mass}, and \Cref{def:heavy-light} give
\begin{align*}
 (3\rho/8)N_k^{H}&\le\sum_{v\in H}b_vx_v\le B.
\end{align*}
For light birth rows, \Cref{lem:helper} and disjointness give
\begin{align*}
 (\alpha\rho/8)N_k^{L}
   &\le\sum_{v\in S\cap V_{\mathrm{light}}}b_v\le4\alpha B.
\end{align*}
Adding these inequalities and using $\tau_k=16\rho$ proves
$\tau_kN_k\le16(8/3+32)B=(1664/3)B$.
\end{proof}

\section{Buying through high girth, and actual physical rental}
\label{sec:bounds}

\subsection{Auxiliary graphs use physical vertices}

Let $H_k$ be the auxiliary graph of edges recorded at scale
$\tau_k$, with vertex set consisting exactly of the endpoints of those
recorded edges. These are physical original endpoint centers, not separate
copies of their birth records; unused isolated centers are omitted. Put
$N_{\ge k}=\sum_{j\ge k}N_j$.

\begin{figure}[t]
\centering
\def\dem{Red!70}
\def\demi{Blue!70}
\def\col{Black!70}
\begin{subfigure}[b]{0.47\textwidth}
\centering
\begin{tikzpicture}[>=Stealth,scale=.95,
  term/.style={circle,draw=\col,fill=\col,line width=1.0pt,inner sep=1.8pt},
  ctr/.style={circle,draw=\col,fill=white,line width=1.0pt,inner sep=2.2pt}]
\path (0,-1.65) rectangle (0,1.65);
\draw[\demi,dashed,line width=1.0pt] (1.2,0) circle (1.6cm);
\draw[\dem,dashed,line width=1.0pt] (3.6,0) circle (1.6cm);
\node[ctr,draw=\demi,label={[text=\col]below:$s$}] (sb) at (1.2,0) {};
\node[term,label={[text=\col]below:$v$}] (vb) at (2.4,0) {};
\node[ctr,draw=\dem,label={[text=\col]below:$s'$}] (spb) at (3.6,0) {};
\draw[\col,line width=1.4pt] (sb)--(vb)--(spb);
\end{tikzpicture}
\caption{Disjoint toll balls at scale $\tau_k$.}
\label{fig:buying-a}
\end{subfigure}\hfill
\begin{subfigure}[b]{0.49\textwidth}
\centering
\begin{tikzpicture}[>=Stealth,scale=.95,
  ctr/.style={circle,draw=\col,fill=white,line width=1.0pt,inner sep=2.2pt}]
\path (0,-1.65) rectangle (0,1.65);
\node[ctr,label={[text=\col]below:$u$}] (ug) at (0,-0.45) {};
\node[ctr] (w1) at (1.8,0.65) {};
\node[ctr] (w2) at (4.4,0.65) {};
\node[ctr,label={[text=\col]below:$v$}] (vg) at (6.2,-0.45) {};
\draw[\demi,line width=1.6pt] (ug)--(w1)--(w2)--(vg);
\node[font=\small,above,text=\col] at (3.1,0.7) {bypass};
\draw[\dem,dashed,line width=1.4pt] (ug)--node[below,font=\small,text=\col] {$uv$} (vg);
\end{tikzpicture}
\caption{High-girth bypass in $H_k$.}
\label{fig:buying-b}
\end{subfigure}
\caption{Two-part buying analysis. (\subref{fig:buying-a}) If two birth balls of radius $\rho=\tau_k/16$ shared a bought node $v\in S$, concatenating their arms through $v$ would connect $s'$ to the earlier center $s$ within mixed cost $2\rho=\tau_k/8$. (\subref{fig:buying-b}) In the auxiliary graph $H_k$, a path of at most $2r-1$ earlier edges between $u$ and $v$ already forms a bought bypass shorter than $QL_i$, preventing any cycle of length at most $2r$.}
\label{fig:buying}
\end{figure}
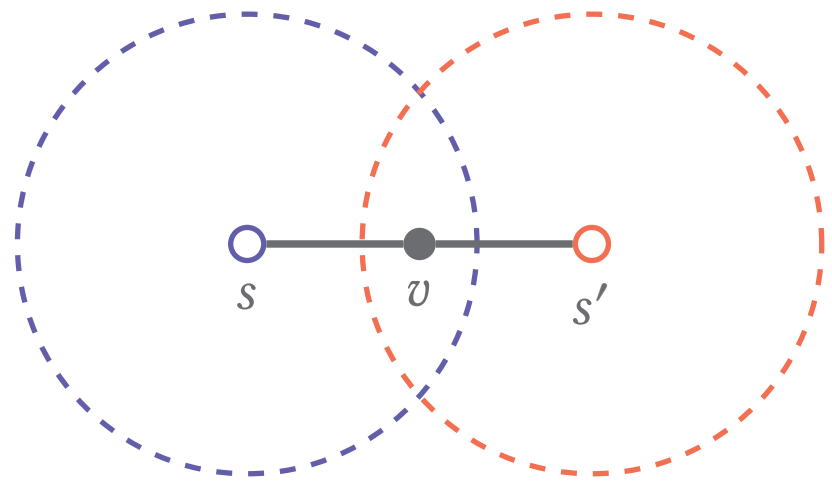
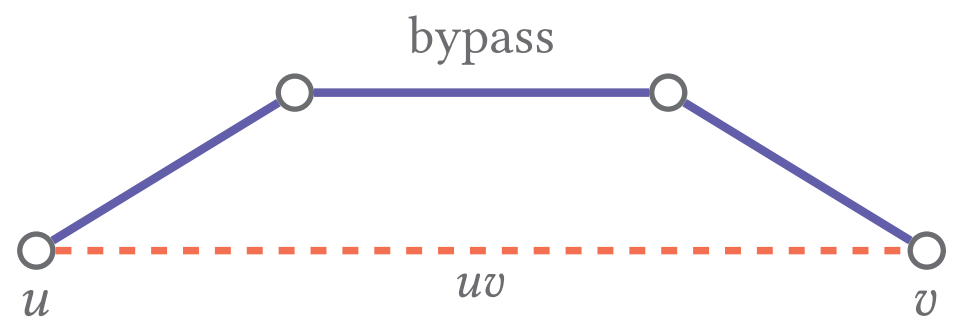

\begin{lemma}[Auxiliary vertices and girth]\label{lem:aux-girth}
The auxiliary graph $H_k$ has girth greater than $2r$ and satisfies
\begin{align*}
 |V(H_k)|&\le\min\{2h,N_{\ge k}\}.
\end{align*}
\end{lemma}
\begin{proof}
Every endpoint of an edge in $H_k$ has a birth record at a scale at least $\tau_k$. By \Cref{lem:attachment}, there are at most $2h$ records in total. This proves the vertex bound.

An endpoint born or promoted at a scale was not previously eligible there. It consequently had no older incident auxiliary edge at that scale. An edge inserted because of a birth cannot be the last edge of a cycle. For any other insertion from request $i$, its endpoints had prior bought physical distance greater than $QL_i$. Every earlier edge on a possible cycle has a bought realization shorter than $3L_i$, by \Cref{lem:attachment} and the nondecreasing processing order of allowances. A cycle of at most $2r$ edges would therefore give a previous path shorter than $3(2r-1)L_i<QL_i$ (\Cref{fig:buying-b}), a contradiction. In particular, there are no loops or parallel edges.
\end{proof}

\begin{lemma}[Elementary girth bound]\label{lem:girth}
An $n$-vertex graph of girth greater than $2r$, $r\ge1$, has
at most $n(1+n^{1/r})$ edges.
\end{lemma}
\begin{proof}
Let $m$ be the number of edges. The assertion is immediate if $m\le n$. Otherwise set
$a=m/n>1$. Repeatedly delete vertices of current degree less than
$a$. This cannot delete every vertex, since their removal degrees
would sum to less than $na=m$. A remaining subgraph has minimum
degree at least $a$. Breadth-first exploration from one of its
vertices to depth $r$ has at least $(a-1)^r$ distinct vertices:
a collision between two branches before that depth would give a
cycle of length at most $2r$. Thus $(a-1)^r\le n$, proving
$m\le n(1+n^{1/r})$.
\end{proof}

\begin{lemma}[Auxiliary edge count]\label{cor:aux-edges}
With $U=1+(2h)^{1/r}$, one has
\begin{align*}
 |E(H_k)|&\le U N_{\ge k}.
\end{align*}
\end{lemma}
\begin{proof}
Apply \Cref{lem:girth} to \Cref{lem:aux-girth} and use $|V(H_k)|\le2h$. Intersections of physical realizing paths only add available routes, so they preserve the last-edge contradiction in \Cref{lem:aux-girth}.
\end{proof}

\subsection{Bounding buying cost}

\begin{lemma}[Buying in one run]\label{lem:buying}
Excluding the supplied seed graph, a run adds buying cost at most $O(J+\lambda K[1+(2h)^{1/r}])B$.
\end{lemma}
\begin{proof}
Each non-base, non-completion buying event records exactly one
auxiliary edge and costs at most $(\lambda+1/4)\tau_k$ by \Cref{lem:attachment}.
Exchanging finite sums and using the common $\lambda$-grid gives
\begin{align*}
 C_{\mathrm{aux}}
 &\le (\lambda+1/4)U\sum_k\tau_kN_{\ge k}\\
 &\le (\lambda+1/4)U\frac{\lambda}{\lambda-1}
                          \sum_j\tau_jN_j\\
 &\le C_0(\lambda+1/4)\frac{\lambda}{\lambda-1}
                            K[1+(2h)^{1/r}]B.
\end{align*}
The last step uses \Cref{lem:packing} and the at most $K$
birth scales above $\eta$. Exceptional completions multiply
within-request non-completion buying by at most $16/15$, by \Cref{lem:attachment}.
Including the helper (\Cref{lem:helper}) and the cheap purchases (\Cref{lem:attachment}), one run therefore adds at most
\begin{align*}
 &4\alpha B+\frac{16}{15}\left[
 B+C_0(\lambda+1/4)\frac{\lambda}{\lambda-1}
                 K[1+(2h)^{1/r}]B\right],
\end{align*}
which is $O(J+\lambda K[1+(2h)^{1/r}])B$ since $\alpha=O(J)$ and $\lambda/(\lambda-1)\le 8/7$. This bound excludes any seed graph supplied before the run.
\end{proof}

\subsection{Bounding physical rental}

\begin{lemma}[Length of an actual route]\label{lem:rental}
The final bought graph contains, for every request $i$ in the run, a route of physical length at most $[(2K+1)(6r+6)+1/4]L_i$.
\end{lemma}
\begin{proof}
Every attachment chain has at most $K$ skipped levels. Each skipped level appends an actual bought route of length at most $QL_i$. The chain ends with a terminal arm of length at most $L_i/8$. The main middle
route adds at most $QL_i$. Thus the resulting route for every
request has physical length at most
\begin{align*}
 &\left[(2K+1)(6r+6)+\frac14\right]L_i.
\end{align*}
All direct and exceptional branches are shorter. Retained paths
remain available as further resources are bought. Replacing the
final route by a shortest physical path only improves the bound.
No purchased resource has been treated as having zero rental.
\end{proof}

\section{Zero-rental preprocessing and separate budgets}
\label{sec:separate}

\begin{lemma}[Zero buying budget]\label{lem:zero-buying}
A feasible LP solution with $B=0$ can be rounded with zero buying and rental at most $R$.
\end{lemma}
\begin{proof}
If $B=0$, all positive LP flow uses zero-buying resources. Buying
that zero-buying support and using shortest physical paths gives
buying zero and rental at most $R$.
\end{proof}

Assume henceforth that $B>0$.

\begin{lemma}[Direct zero-length rounding]
\label{lem:zero}
For any subset of requests whose unit LP flows are supported on
zero-physical-length paths, one can connect those requests at
buying cost $O(\log(2h))B$ and physical rental zero.
\end{lemma}
\begin{proof}
If the selected subset is empty, buy nothing. Otherwise $h\ge1$ and
$J=\lceil\log_2(2h)\rceil\ge1$. Restrict the graph to zero-physical-length
resources, retaining the selected requests and their unit LP flows. Their
terminal leaves remain in this graph by \Cref{lem:node-reduction}. Retain the
openings of the remaining resources and let $B_0\le B$ be their fractional
buying cost. The restricted solution has rental zero. If $B_0=0$, apply
\Cref{lem:zero-buying} to this solution, obtaining buying and rental zero.

Suppose $B_0>0$. Run the ordinary core algorithm with empty seed,
$\lambda=8$, $r=J$, and $L_i=1$ for every selected request; thus $Q=6J+6$.
Keep the original $h$ as the normalization in the grid, cutoff, and helper.
The required allowance condition holds since $\ell(f_i)=0\le L_i/64$.
Moreover,
\begin{align*}
 K=\left\lceil\log_8(16h)\right\rceil\le J+1,
 \qquad (2h)^{1/J}\le2.
\end{align*}
The earlier one-run buying bound, \Cref{lem:buying}, therefore gives
\begin{align*}
 b(F)&=O\!\left(J+8K[1+(2h)^{1/J}]\right)B_0
       =O(J)B_0=O(\log(2h))B.
\end{align*}
Every selected route lies in the restricted graph and hence has physical
length zero. This proves the rental assertion directly, without charging
an artificial rental budget for the allowances $L_i=1$.
\end{proof}

\begin{theorem}[Parameterized rounding guarantee]
\label{thm:main}
For integers $\lambda\ge8$ and $r\ge1$, put
\begin{align*}
 K&=\left\lceil\log_\lambda(2\lambda h)\right\rceil,
 \qquad J=\lceil\log_2(2h)\rceil.
\end{align*}
Every feasible rational solution of the natural affine LP (\Cref{def:natural-lp}) can be rounded
deterministically in polynomial time to a bought graph $F$ satisfying
\begin{align*}
 b(F)&=O\!\left(J+\lambda K[1+(2h)^{1/r}]\right)B,
 \qquad
 \sum_{i=1}^h\delta_i d_{F,\ell}(s_i,t_i)=O(rK)R.
\end{align*}
Constants are absolute and independent of the numerical demand ratio.
The assertion holds simultaneously for node and edge costs.
\end{theorem}
\begin{proof}
If $B=0$, apply \Cref{lem:zero-buying}; its estimates imply the theorem. Assume $B>0$.
For a positive-demand request, $R_i=0$ implies that every positive
LP path has zero physical length. If there are such requests,
apply \Cref{lem:zero}. If they are all requests, stop.
Otherwise retain the resulting bought graph but discard all its
center and auxiliary records. For the remaining requests set
\begin{align*}
 L_i&=64R_i/\delta_i>0.
\end{align*}
The original $h$ remains the normalization in the grid, cutoff,
helper, and theorem, even if fewer requests remain. Their original
LP is still feasible after the seed purchases. In particular,
\begin{align*}
 \ell(f_i)&=L_i/64,\qquad \sum_{i:R_i>0}\delta_iL_i=64R.
\end{align*}
Run the algorithm of \Cref{sec:algorithm}, with the helper of \Cref{sec:helper}, on those requests.
Combining \Cref{lem:buying} with the zero-length buying cost (\Cref{lem:zero}) gives the buying bound in \Cref{thm:main}. Combining \Cref{lem:rental} with $\sum_{i:R_i>0}\delta_iL_i=64R$ gives the explicit rental
factor
\begin{align*}
 64\left[(2K+1)(6r+6)+\frac14\right]R&=O(rK)R.
\end{align*}
This completes the proof of \Cref{thm:main} without charging
zero-rental requests a positive artificial rental budget.
\end{proof}

\section{Optimization and algorithmic details}\label{sec:optimization}

\begin{lemma}[Optimized parameter bounds]\label{lem:param-bounds}
Choose the integers
\begin{align*}
 q&=\min\{j\in\mathbb Z:j\ge2,\ j^j\ge2h\},\qquad
 r=2q,\qquad \lambda=\max\{8,\lceil\sqrt q\rceil\}.
\end{align*}
Then $K=\lceil\log_\lambda(2\lambda h)\rceil$ and $J=\lceil\log_2(2h)\rceil$ satisfy
\begin{align*}
 K&\le2q+1,\qquad J=O(q^2),\qquad
 \lambda K[1+(2h)^{1/r}]=O(q^2),\qquad rK=O(q^2),
\end{align*}
where $q=O(\log(2h)/\log\log(4h))$ (so $q=O(\log h/\log\log h)$ for $h\ge3$, and $q=O(1)$ for $h\le2$).
\end{lemma}
\begin{proof}
Since $2h\le q^q$, we have
\begin{align*}
 (2h)^{1/r}&\le\sqrt q,\qquad \lambda=O(\sqrt q),
 \qquad \lambda^{2q}\ge q^q\ge2h.
\end{align*}
Consequently $K=1+\lceil\log_\lambda(2h)\rceil\le2q+1$, and both $\lambda K[1+(2h)^{1/r}]$ and $rK$ are $O(q^2)$.
Since $2h\le q^q$, the additive $J$ term also satisfies
\begin{align*}
 J&=\lceil\log_2(2h)\rceil
  \le q\log_2 q+1=O(q^2).
\end{align*}
Finally, put $T=\ln(2h)$. For sufficiently large $T$, the integer $j=\lceil2T/\ln T\rceil$ satisfies $\ln j\ge(\ln T)/2$, hence $j\ln j\ge T$. By minimality, $q\le j$; the remaining bounded values of $h$ change only the absolute constant. Thus $q=O(\log(2h)/\log\log(4h))$, which is $O(\log h/\log\log h)$ for $h\ge3$ and bounded for $h\le2$.
\end{proof}

\begin{theorem}[Optimized rounding of the affine LP]\label{thm:lp-rounding}
There is a deterministic polynomial-time algorithm that rounds any feasible rational solution of the natural affine LP (\Cref{def:natural-lp}) with $h\ge3$ demand pairs and buying and rental costs $B,R$ to a bought graph $F$ such that
\begin{align*}
 b(F)&\le C\,\frac{\log^2 h}{(\log\log h)^2}\,B,\\
 \sum_{i=1}^h\delta_i d_{F,\ell}(s_i,t_i)
 &\le C\,\frac{\log^2 h}{(\log\log h)^2}\,R,
\end{align*}
where $C$ is an absolute constant. The guarantee holds for arbitrary positive rational demands and for buying prices and physical lengths on nodes and edges simultaneously.

For one or two pairs, both estimates hold with an absolute constant factor in place of the displayed logarithmic factor.
\end{theorem}
\begin{proof}
Substitute the parameter choices and bounds of \Cref{lem:param-bounds} into \Cref{thm:main}, together with the polynomial-time implementation in \Cref{lem:runtime}.
\end{proof}

\begin{lemma}[Polynomial implementation]\label{lem:runtime}
For binary-encoded integers $\lambda\ge8$ and $r\ge1$, the parameterized algorithm uses polynomially many rational arithmetic operations with polynomial encoding length in the graph, supplied LP solution, and parameter encodings. In particular, the optimized algorithm has this property. Its rounding stages have no dependence on the numerical demand ratio beyond the input encoding length.
\end{lemma}
\begin{proof}
All parameter choices can use exact integer comparisons. In
particular, compute $q$ using $j^j\ge2h$, compute the integer
ceiling of $\sqrt q$, and compute $K$ by comparing integer powers
of $\lambda$. No floating-point logarithm is needed to select a
scale or make a feasibility decision.
For arbitrary supplied parameters, $K\le J+1=O(\log(2h))$ and $r$
enters the implementation only through the scalar $Q=6r+6$.
The powers of $\lambda$ used in the grid have exponents of magnitude
at most $K$, so their binary lengths are $O(K\log\lambda)$; the
binary length of $Q$ is $O(1+\log r)$. Thus these choices do not
introduce loops polynomial in the numerical magnitudes of the parameters.

There are $O(h\log(2h))$ fixed helper rows and at most
$O(h\log(2h))$ recursion invocations. Mixed and bought-graph
distances are ordinary nonnegative shortest-path computations;
there is no length-constrained shortest-path oracle. Products in
the conditional-expectation method have polynomially many rational
factors. Each mixed shortest-path weight is formed from fixed input
prices, a grid scale, and a fixed request allowance; each distance is a
sum along a simple path. These scalars have polynomial binary length.
Each helper potential is a sum of polynomially many products of fixed
rational factors and chosen zero--one indicators, so its numerator and
denominator lengths are polynomial as well. The common grid spans a
ratio less than $2\lambda^2h$, and every recursive birth is above its cutoff. The running time and
approximation ratio have no dependence on the magnitude of the
demand ratio beyond the ordinary encoding length of the input.
\end{proof}

\begin{proof}[Proof of \Cref{thm:rounding}]
By \Cref{lem:node-reduction,lem:lp-solution}, reduce the input instance to an equivalent node-cost instance and solve and decompose the natural affine LP (\Cref{def:natural-lp}) in polynomial time. Its fractional optimum $B+R$ is at most the minimum integral routing cost, since every integral routing is a feasible LP solution. Apply \Cref{thm:lp-rounding} and add its separate buying and rental estimates to bound the integral objective by $O(\log^2 h/(\log\log h)^2)(B+R)$ for $h\ge3$ (and $O(B+R)$ for $h\le2$). This proves both the integrality-gap upper bound and the stated approximation factor; \Cref{lem:runtime} gives a deterministic polynomial implementation, and \Cref{lem:node-reduction} maps the rounded solution back to the original instance while preserving both buying and rental costs.
\end{proof}

\section{A near-logarithmic integrality gap for edge costs}
\label{sec:edge-lower}

This section proves \Cref{thm:edge-gap}. The node-cost construction of \Cref{sec:node-lower} relies on node buying prices, which edge-only instances do not have. We therefore give a separate construction in which only edges carry buying prices and physical lengths. It combines a high-girth regular demand graph, a polynomial-size covering lift with few cycles in small subgraphs, and a mod-$2$ dimension bound on requests routed along short connections. The result concerns the natural LP only; it is not a hardness-of-approximation result.

Throughout this section, every vertex has zero buying price and zero physical length, and all demands are one. By \Cref{lem:node-reduction}, the natural affine LP (\Cref{def:natural-lp}) of such an instance is equivalent to the edge formulation with opening variables $x_e\in[0,1]$ and path-flow variables $f_{iP}\ge0$ satisfying
\begin{align*}
 \sum_{P\in\mathcal P_i}f_{iP}&=1\quad(i\in[h]),\qquad
 f_i(e):=\sum_{P\in\mathcal P_i:e\in P}f_{iP}\le x_e\quad(i\in[h],\ e\in E),
\end{align*}
and objective $\sum_{e\in E}b_ex_e+\sum_{i=1}^h\sum_{P\in\mathcal P_i}\ell(P)f_{iP}$. As in \Cref{sec:node-lower}, we write $\LP$ and $\OPT$ for the fractional and integral optima.

\subsection{Cycle rank and lifts}
For a graph $W$, let $V_+(W)$ be its set of nonisolated vertices and $c(W)$ the number of its components that contain an edge. Its \emph{cycle rank}
\begin{align*}
 \beta(W)&:=|E(W)|-|V_+(W)|+c(W)
\end{align*}
is the dimension of its cycle space over $\mathbb F_2$, and $\beta(\varnothing)=0$. The cycle rank is additive over components.

Let $\Gamma$ be a simple graph. For an integer $p\ge1$, a \emph{$p$-lift} $\widetilde\Gamma$ of $\Gamma$ has one vertex $(v,j)$ for every $v\in V(\Gamma)$ and $j\in[p]$; for each base edge $uv\in E(\Gamma)$, its $p$ lifted edges form a perfect matching between the \emph{fibers} $\{u\}\times[p]$ and $\{v\}\times[p]$. In a \emph{random} $p$-lift~\cite{AL02}, these matchings are independent and uniformly random. The projection $(v,j)\mapsto v$ maps each lifted edge to its base edge. Every vertex $(v,j)$ is incident to exactly one lifted edge over each base edge at $v$, so a cycle of $\widetilde\Gamma$ projects to a closed nonbacktracking walk of the same length in $\Gamma$, which contains a cycle of no greater length. Hence a lift of a graph of girth at least $g$ also has girth at least $g$.

\begin{lemma}[Local cycle sparsity of a lift]\label{lem:lift}
Let $\Gamma$ be a simple graph with $N$ vertices, maximum degree $\Delta\ge4$, and girth at least $g\ge4$. For an integer $T\ge g$, put
\begin{align*}
 A&=64\Delta^2,\qquad p=\left\lceil1000\,NT^2A^g\right\rceil.
\end{align*}
There is a $p$-lift $\widetilde\Gamma$ of $\Gamma$ such that every subgraph $W\subseteq\widetilde\Gamma$ with $|E(W)|\le T$ satisfies
\begin{align*}
 \beta(W)&\le\frac{2|E(W)|}{g}.
\end{align*}
\end{lemma}
\begin{proof}
We show that a random $p$-lift has this property with positive probability. Since the cycle rank and the edge count are additive over components, a violating subgraph has a connected violating component. A connected subgraph with $m<g$ edges is acyclic by the girth of the lift, so it suffices to exclude connected subgraphs $W$ with $g\le m\le T$ edges and $\beta(W)>2m/g$.

Fix such a witness $W$ and a root vertex in it. A depth-first search of $W$ from the root can be encoded by a \emph{script}: at the current vertex $(v,j)$, the next step either returns to the parent, or queries the unexposed lifted edge of $W$ at $(v,j)$ over a specified base edge at $v$, together with a bit predicting whether its other endpoint is new or already discovered. A new endpoint becomes the current vertex. Every edge of $W$ is queried exactly once, and there are at most $m$ returns, so a script is a word of length at most $2m$ over an alphabet of at most $2\Delta+1$ symbols. There are at most $(2\Delta+2)^{2m}\le(4\Delta)^{2m}\le A^m$ such words. A script \emph{succeeds} in a lift if executing it from the root only queries unexposed lifted edges and every query matches its predicted bit. The script of $W$ succeeds, and its number of old predictions is $m-(|V(W)|-1)=\beta(W)$.

Condition on the outcomes of earlier queries of a fixed script. At most $m-1<T$ lifted edges are exposed, so the queried endpoint is uniform among at least $p-T$ unmatched vertices of the target fiber, of which at most $m+1\le T+1$ are discovered. Since $p\ge8T$, the probability that the query meets a discovered vertex is at most $(T+1)/(p-T)\le4T/p$. This bound remains valid when the base edges in the script are chosen adaptively from earlier outcomes. Multiplying over the $\beta(W)>2m/g$ old predictions, a fixed script with a violating number of old predictions succeeds with probability at most $(4T/p)^{2m/g}$.

A union bound over the $Np$ roots, the sizes $m\in[g,T]$, and the scripts bounds the probability that a violating witness exists by
\begin{align*}
 Np\sum_{m=g}^{T}A^m\left(\frac{4T}{p}\right)^{2m/g}
 &\le Np\,\frac{\rho^g}{1-\rho},\qquad
 \rho:=A\left(\frac{4T}{p}\right)^{2/g}.
\end{align*}
Since $p\ge4TA^g$, we have $(4T/p)^{2/g}\le A^{-2}$, and hence $\rho\le1/A<1/2$. Moreover, $\rho^g=A^g(4T/p)^2$, so the probability is at most
\begin{align*}
 2Np\,A^g\left(\frac{4T}{p}\right)^2
 &=\frac{32NT^2A^g}{p}\le\frac{32}{1000}<1.
\end{align*}
Thus a lift with the stated property exists.
\end{proof}

\subsection{The instance and its fractional solution}
Fix an integer $g\ge4$. By the Erd\H{o}s--Sachs theorem~\cite{ES63}, there is a simple $g$-regular graph $\Gamma$ of girth at least $g$ on $N\le8g^g$ vertices. Breadth-first search to depth $\lfloor(g-1)/2\rfloor$ from any vertex of $\Gamma$ explores a tree, which gives the Moore bound
\begin{align*}
 N&\ge(g-1)^{\lfloor(g-1)/2\rfloor}.
\end{align*}
Put
\begin{align*}
 h&=|E(\Gamma)|=\frac{Ng}{2},\qquad T=hg,
\end{align*}
and let $\widetilde\Gamma$ be the $p$-lift of $\Gamma$ given by \Cref{lem:lift} with $\Delta=g$. The supply graph $G$ consists of $\widetilde\Gamma$, whose edges we call \emph{core edges}, together with a \emph{hub} $v^*$ for every $v\in V(\Gamma)$ joined by \emph{spokes} to all $p$ vertices $(v,j)$ of its fiber. Every spoke has buying price one and physical length zero, and every core edge has buying price zero and physical length $2/g$. For each base edge $e=uv\in E(\Gamma)$, there is one unit-demand request $(u^*,v^*)$, for $h$ requests in total.

\begin{lemma}[Fractional solution]\label{lem:edge-lp}
The natural affine LP of this instance has a feasible solution with buying cost $B=N$ and rental cost $R=N$. In particular, $\LP\le2N$.
\end{lemma}
\begin{proof}
Open every spoke and every core edge to extent $1/p$. For the request of a base edge $e=uv$, let $\pi_e$ be the bijection of $[p]$ given by the lifted matching of $e$, and send $1/p$ units of flow along each of the $p$ paths
\begin{align*}
 u^*\,\text{--}\,(u,j)\,\text{--}\,(v,\pi_e(j))\,\text{--}\,v^*,\qquad j\in[p].
\end{align*}
These paths are edge-disjoint, so every edge carries at most $1/p$ units of this request's flow, as required by the per-request opening constraints. The $Np$ spokes contribute buying cost $Np\cdot(1/p)=N$, and core edges are free. Every path has physical length $2/g$, so the rental cost is $h\cdot2/g=N$.
\end{proof}

\subsection{Short connections and the integral optimum}
Consider any integral solution with routes $P_1,\ldots,P_h$. Let $S$ be the number of bought spokes, and let $L$ be the total number of core edges on the routes, counted with multiplicity over the requests. Since spokes have zero length and core edges are free,
\begin{align*}
 b\!\left(\bigcup_{i=1}^hP_i\right)+\sum_{i=1}^h\ell(P_i)&=S+\frac{2L}{g}.
\end{align*}
Let $W\subseteq\widetilde\Gamma$ consist of the core edges used by at least one route, so that $|E(W)|\le L$. Call a request \emph{short} if its route uses at most $g-2$ core edges, and let $h_{\mathrm{short}}$ be the number of short requests.

\begin{lemma}[Short-request dimension bound]\label{lem:short-requests}
Every integral routing satisfies
\begin{align*}
 h_{\mathrm{short}}&\le S+\beta(W).
\end{align*}
\end{lemma}
\begin{proof}
Let $F$ be the graph formed by the $S$ bought spokes and the edges of $W$, and let $Z\subseteq\mathbb F_2^{E(F)}$ be the space of edge sets of $F$ whose boundary vanishes at every lifted vertex $(v,j)$; their boundary may be nonzero at the hubs. The incidence matrix between lifted vertices and the edges of $W$ has rank $|V_+(W)|-c(W)$, witnessed by a spanning forest, and adding the spoke columns cannot decrease this rank. Hence
\begin{align*}
 \dim Z&\le S+|E(W)|-\bigl(|V_+(W)|-c(W)\bigr)=S+\beta(W).
\end{align*}

Define the linear map $\phi:\mathbb F_2^{E(F)}\to\mathbb F_2^{E(\Gamma)}$ that sends every core edge to its base edge and every spoke to zero. It commutes with taking boundaries once each hub $v^*$ and each lifted vertex $(v,j)$ are mapped to $v$; spokes contribute nothing, since both endpoints of a spoke map to the same vertex.

Let $e=uv$ be a short request, and let $z_e\in Z$ be the edge set of its route, a simple $u^*$--$v^*$ path. Its boundary is $u^*+v^*$, so $\phi(z_e)$ has boundary $u+v$ in $\Gamma$. Therefore $\phi(z_e)+\mathbf 1_e$ lies in the cycle space of $\Gamma$, and its support has at most $(g-2)+1<g$ edges. A nonzero element of the cycle space contains a cycle, and $\Gamma$ has girth at least $g$. Consequently, $\phi(z_e)=\mathbf 1_e$. These standard basis vectors are linearly independent for distinct short requests, so $h_{\mathrm{short}}\le\dim\phi(Z)\le\dim Z$.
\end{proof}

\begin{lemma}[Integral lower bound]\label{lem:edge-opt}
Every integral solution of this instance costs at least $Ng/4$. In particular, $\OPT\ge Ng/4$.
\end{lemma}
\begin{proof}
If $L>T=hg$, the rental cost alone exceeds $2hg/g=Ng$. Assume $L\le T$. Every request that is not short uses at least $g-1$ core edges, so $h-h_{\mathrm{short}}\le L/(g-1)$. Since $|E(W)|\le L\le T$, \Cref{lem:lift} gives $\beta(W)\le2L/g$, and \Cref{lem:short-requests} yields
\begin{align*}
 h&\le S+\left(\frac1{g-1}+\frac2g\right)L\le S+\frac{4L}{g},
\end{align*}
where the last step uses $g\ge4$. Therefore,
\begin{align*}
 S+\frac{2L}{g}&\ge\frac12\left(S+\frac{4L}{g}\right)\ge\frac h2=\frac{Ng}{4}.
\end{align*}
\end{proof}

\subsection{Proof of the integrality gap}
\begin{proof}[Proof of \Cref{thm:edge-gap}]
For every integer $g\ge4$, take the instance above. By \Cref{lem:edge-lp,lem:edge-opt},
\begin{align*}
 \frac{\OPT}{\LP}&\ge\frac{Ng/4}{2N}=\frac g8.
\end{align*}
The upper bound $N\le8g^g$ and the Moore bound give $\log h=\log(Ng/2)=\Theta(g\log g)$, and hence $g=\Theta(\log h/\log\log h)$. Since $h\to\infty$ as $g\to\infty$, this gives infinitely many instances with integrality gap $\Omega(\log h/\log\log h)$.

For the size bound, the Moore bound gives $g\log g=O(\log N)$, and $A=64g^2\le g^5$ for $g\ge4$, so $A^g=N^{O(1)}$. Since $N\le h$ and $T=hg$, the lift order $p$ is polynomial in $h$. The supply graph has $N(p+1)$ vertices and $p(h+N)$ edges.

It remains to make all edge data strictly positive and the terminal pairs disjoint. Give every core edge buying price $1/h$ instead of zero, and every spoke physical length $1/h$ instead of zero. For every request $e=uv$, attach private degree-one leaves $s_e$ and $t_e$ to $u^*$ and $v^*$ by \emph{pendant} edges of buying price and physical length $1/(10h)$, and replace the request by $(s_e,t_e)$. Every route of the new request consists of its two pendant edges and a $u^*$--$v^*$ route in $G$, since a simple path cannot pass through another degree-one leaf. Hence, no integral cost decreases, and \Cref{lem:edge-opt} still applies. For the fractional solution of \Cref{lem:edge-lp}, extended by opening every pendant edge to extent one, the core edges add buying cost $hp\cdot(1/p)\cdot(1/h)=1$, the spokes add rental cost $h\cdot2/h=2$, and the pendant edges add at most $4h/(10h)<1$. The fractional objective is therefore less than $2N+4\le3N$, because $N\ge g+1\ge5$, and the ratio is at least $g/12$. All buying prices and physical lengths are positive rationals with $O(\log h)$-bit encodings, all vertices have zero buying price and zero physical length, the terminal pairs are pairwise vertex-disjoint, and the supply graph remains polynomial in $h$.
\end{proof}

Together with \Cref{thm:rounding}, \Cref{thm:edge-gap} shows that the worst-case integrality gap of the natural affine LP for edge-only instances is $\Omega(\log h/\log\log h)$ for infinitely many $h$ and $O(\log^2h/(\log\log h)^2)$ for all $h\ge3$.

\section*{AI Acknowledgement}
Some of the proofs in this paper were developed with the help of OpenAI's GPT Astra, which also contributed to the writing and exposition, with Gemini Argon and Claude Opus assisting in some parts. The authors are responsible for the correctness of the results and the final content of the paper. Lean files formalizing the main theorem are available on GitHub at \url{https://github.com/igholami/buy-at-bulk-lean}.

\bibliographystyle{alpha}
\bibliography{references}

\appendix
\section{A logarithmic integrality gap for node costs}
\label{sec:node-lower}

For completeness, we verify the classical node-cost LP integrality gap mentioned in \Cref{sec:related}. The construction is the standard set-cover example using all $k$-subsets of a $2k$-element ground set (see Vondr{\'a}k~\cite{Vondrak05}), realized via the usual set-cover reduction to node-weighted Steiner tree~\cite{KR95}.

\begin{proposition}\label{thm:node-lower}
For every integer $h\ge2$, there is an instance with $h$ requests whose natural node-LP (\Cref{def:natural-lp}) integrality gap is greater than $\frac14\log_2 h$. All physical lengths can be zero.
\end{proposition}
\begin{proof}
Fix an integer $k\ge1$, and let
$\mathcal A_k=\{A\subseteq[2k]:|A|=k\}$ and
$h_k=|\mathcal A_k|=\binom{2k}{k}$.
Create a root $s$, nodes $v_1,\ldots,v_{2k}$, and a terminal $t_A$
for every $A\in\mathcal A_k$. The edges are
\begin{align*}
 &\{sv_j:j\in[2k]\}
 \;\cup\;
 \{v_jt_A:A\in\mathcal A_k,\ j\in A\}.
\end{align*}
Each $v_j$ has buying cost one. The root and all terminals have
buying cost zero; all edge buying costs and all physical lengths
are zero. There is one request $(s,t_A)$ of demand one for each
$A\in\mathcal A_k$. We first show that
\begin{align*}
 \LP&=2,\qquad \OPT=k+1.
 \numberthis\label{eq:node-gap-optima}
\end{align*}

To upper-bound the fractional optimum, set $x_{v_j}=1/k$ for every $j$, and set the opening variables of
all free resources to one. For request $A$, send $1/k$ units of
flow on each path $(s,v_j,t_A)$ with $j\in A$. These $k$ paths
carry one unit in total and use every paid node to extent at most
$1/k$, as required by the per-commodity opening constraints.
The buying cost is $2k/k=2$, and the rental cost is zero.

For the reverse inequality, consider any feasible solution of the
node relaxation in \Cref{sec:prelim}. Every path ending at $t_A$ contains
some $v_j$ with $j\in A$, namely its last neighbor of $t_A$.
Consequently,
\begin{align*}
 \sum_{j\in A}x_{v_j}
 &\ge
 \sum_{j\in A}\sum_{\substack{P\in\mathcal P_A\\v_j\in P}}f_{AP}
 =
 \sum_{P\in\mathcal P_A}f_{AP}
       \bigl|\{j\in A:v_j\in P\}\bigr|
 \ge1.
 \numberthis\label{eq:node-gap-cover}
\end{align*}
Here $\mathcal P_A$ denotes all simple $s$--$t_A$ paths. In
particular, the inequality also accounts for paths that pass
through other terminals. Summing~\Cref{eq:node-gap-cover} over
$A\in\mathcal A_k$ and using the fact that each index belongs to
$\binom{2k-1}{k-1}=h_k/2$ such sets gives
\begin{align*}
 \frac{h_k}{2}\sum_{j=1}^{2k}x_{v_j}&\ge h_k.
\end{align*}
Thus every feasible LP solution has buying cost at least two,
proving $\LP=2$.

For the integral optimum, let $C=\{j:v_j\text{ is bought}\}$ in an integral solution.
Connectivity of $t_A$ requires $C\cap A\ne\varnothing$.
If $|C|\le k$, the complement $[2k]\setminus C$ contains a
$k$-subset $A$, and $t_A$ has no bought neighbor. Hence every
feasible solution costs at least $k+1$. Conversely, any $k+1$
indices meet every $k$-subset. Buying their nodes connects every
terminal to $s$ by a path of two edges. This proves
$\OPT=k+1$ and establishes~\Cref{eq:node-gap-optima}.
Since $h_k\le2^{2k}$, we obtain
\begin{align*}
 \frac{\OPT}{\LP}&=\frac{k+1}{2}
 \ge\frac14\log_2 h_k.
\end{align*}

Finally, for a prescribed $h\ge2$, choose the largest $k$ with $h_k\le h$
and repeat one of the requests until there are exactly $h$.
Repeating a request changes neither optimum: its connectivity
requirement is already present, its LP flow can be copied, and
all rental costs are zero. Since
$h<h_{k+1}\le2^{2k+2}$, the resulting ratio satisfies
\begin{align*}
 \frac{\OPT}{\LP}&=\frac{k+1}{2}>
 \frac14\log_2 h.
\end{align*}
If distinct terminal occurrences are desired, the free endpoint-leaf transformation in \Cref{sec:prelim} preserves both optima.
This completes the proof.
\end{proof}

The example has only node buying costs and already lies in the
single-source case. Thus the logarithmic loss in the zero-rental
preprocessing of \Cref{sec:separate} is necessary for rounding
the natural node LP. More generally, it gives the
$\Omega(\log h)$ lower endpoint of the integrality-gap range stated
in the introduction.

\end{document}